\documentclass[useAMS,usenatbib]{mn2e}

\usepackage{graphics,mssymb}

\title[An X-ray emission line spectrum of Nova V382\,Velorum 1999]{An X-ray emission line spectrum of Nova V382\,Velorum 1999}
\author[J.-U. Ness and S. Starrfield and C. Jordan and J. Krautter and J.H.M.M.
Schmitt]{J.-U. Ness$^{1,2}$\thanks{E-mail: ness@thphys.ox.ac.uk(JUN)}
S. Starrfield$^3$, C. Jordan$^1$, J. Krautter$^4$, and J.H.M.M. Schmitt$^2$\\
$^1$Department of Physics, Rudolf Peierls Centre for Theoretical Physics,
University of Oxford, 1 Keble Road, Oxford OX1\,3NP, UK\\
$^2$Hamburger Sternwarte, Universit\" at Hamburg, Gojenbergsweg 112,
D-21029 Hamburg, Germany\\
$^3$Department of Physics and Astronomy, Arizona State University,
Tempe, AZ 85287-1504, USA\\
$^4$Landessternwarte K\"onigstuhl, D-69117 Heidelberg, Germany\\
}

\begin{document}

\date{Accepted ; Received \today}

\pagerange{\pageref{firstpage}--\pageref{lastpage}} \pubyear{2005}

\maketitle

\label{firstpage}

\begin{abstract}
We report on the analysis of an X-ray grating spectrum of the
Classical Nova V382\,Vel (1999), obtained with the LETG+HRC-S
instrument on board CHANDRA, which shows emission lines dominating over
any continuum. Lines of
Si, Mg, Ne, O, N, and C are identified, but no Fe lines are detected. The total
luminosity in the lines is $\sim 4\times 10^{27}$\,erg\,s$^{-1}$ (corrected for
N$_{\rm H}=1.2\times 10^{21}$\,cm$^{-2}$). The lines have broad profiles with
FWHM corresponding to a velocity $\sim 2900\,\pm\,200$\,km\,s$^{-1}$. Some
structure is identified
in the profiles, but for different elements we find different profile structures.
While lines of O show a broadened Gaussian profile, those of Ne are double-peaked,
suggesting a fragmented emitting plasma. Using the emission measure distribution
we derive relative element abundances and find abundances of Ne and N that are
significantly
enhanced relative to that of O, while Fe is not over-abundant. The lack of any
source emission longwards of 50\,\AA\ and the O\,{\sc viii} Ly$_\alpha$/Ly$_\beta$
line ratio support previous values of the hydrogen column-density. We find weak
continuum emission from the white dwarf, consistent with a black-body spectrum with
an upper limit to the temperature of $T=3\times 10^5$\,K, assuming a source
radius of 6000\,km. The upper limit for the integrated black-body luminosity is
$2\times 10^{36}$\,erg\,s$^{-1}$. The {\it BeppoSAX} and {\it Chandra}
ACIS observations of V382\,Vel show that the nova
was bright and in the Super Soft phase as late as
1999 December 30. Our LETG observation obtained
6 weeks later, as well as all subsequent X-ray observations,
showed a remarkable fading to a nearly pure emission
line phase which suggests that nuclear burning on the
white dwarf had turned off by February. In the absence of a photoionizing source the
emission lines were formed in a collisionally ionized and excited expanding shell.
\end{abstract}


\section{Introduction}

\label{intro}

Classical Novae (CN), a class of Cataclysmic Variables (CVs), are thought
to be thermonuclear explosions induced on the surface
of white dwarfs as a result of continuous accretion of material from a
companion main sequence star. A sufficient accumulation of hydrogen-rich
fuel causes a thermonuclear runaway (TNR). Extensive modelling of the TNR has
been carried out in the past \citep[e.g.,][ and references therein]{st89}.
These models showed that only a part of the ejected envelope
actually escapes, while the remaining material forms an envelope on the white
dwarf with ongoing nuclear burning, radiation driven winds, and turbulent motions.
These processes result in a shrinking of the nuclear burning white dwarf radius with
increasing temperatures \citep{st91,kr02}. During this phase of ``constant bolometric
luminosity" the nova emits strong X-ray radiation with a soft spectral signature.\\
Classical Novae have been observed with past X-ray missions, e.g., {\it Einstein},
{\it ROSAT}, {\it ASCA}, and {\it BeppoSAX}. While X-ray lightcurve variations were
studied, the X-ray spectra obtained had low dispersion and were quite limited. The
transmission and reflection gratings aboard {\it Chandra} and {\it XMM-Newton} now
provide significantly improved sensitivity and spectral resolution, and these
gratings are capable of resolving individual emission or absorption lines. The
{\it Chandra} LETG (Low Energy Transmission Grating) spectrum of the Classical
Nova V4743\,Sgr \citep{v4743} showed strong continuum emission with superimposed
absorption lines, while V1494\,Aql showed both absorption and emission lines
\citep{drake03}. Essentially all X-ray spectra of Classical Novae
differ from each other, so no classification scheme has so far been established.
A review of X-ray observations of novae is given by \cite{orio04}.\\

\begin{table*}
\caption{\label{pobs}Summary of X-ray, UV and optical observations of V382\,Vel}
\begin{tabular}{lcccr}
\hline
Date & day after outburst & Mission & Remarks & Reference\\
\hline
&5--498& La Silla & $V_{\rm max}=2.3$ (1999 23 May)& \cite{vall02}\\
&&&fast Ne Nova; d=1.7\,kpc ($\pm20$ per cent)&\\
1999 May 26& 5.7 & {\it RXTE} & faint in X-rays & \cite{mukai01} \\
1999 June 7& 15 & {\it BeppoSAX} & first X-ray detection & \cite{orio01a}\\
&&&no soft component &\\
1999 June 9/10 & 20.5 & {\it ASCA} & highly absorbed bremsstrahlung & \cite{mukai01} \\
1999 June 20& 31 & {\it RXTE} & decreasing plasma temperature & \cite{mukai01} \\
1999 June 24& 35 & {\it RXTE} & and column-density& \cite{mukai01} \\
1999 July 9& 50 & {\it RXTE} & $\stackrel{.}{.}$ & \cite{mukai01} \\
1999 July 18& 59 & {\it RXTE} & $\stackrel{.}{.}$ & \cite{mukai01} \\
1999 May 31-- & & {\it HST}/STIS & UV lines indicate fragmentation of& \cite{shore03} \\
\ \ 1999 Aug 29& & &ejecta; & \\
 & & &C and Si under-abundant,&\\
 & & &O, N, Ne, Al over-abundant&\\
 & & &$d=2.5$\,kpc; N$_{\rm H}=1.2\times 10^{21}$\,cm$^{-2}$&\\
1999 Nov 23& 185 & {\it BeppoSAX} & hard and soft component & \cite{orio01a,orio02} \\
1999 Dec 30& 223 & {\it Chandra} (ACIS) & & \cite{burw02}\\
2000 Feb 6-- Jul 3 & & {\it FUSE} & O\,{\sc vi} line profile&\cite{shore03}\\
{\bf 2000 Feb 14} & {\bf 268} & {\it Chandra} (LETG) & {\bf details in this paper} &\cite{burw02}\\
2000 Apr 21& 335 & {\it Chandra} (ACIS) & & \cite{burw02}\\
2000 Aug 14& 450 & {\it Chandra} (ACIS) & & \cite{burw02}\\
\hline
\end{tabular}
\end{table*}

\section{The Nova}
\subsection{V382\,Velorum}

\begin{figure*}
\resizebox{\hsize}{!}{\includegraphics{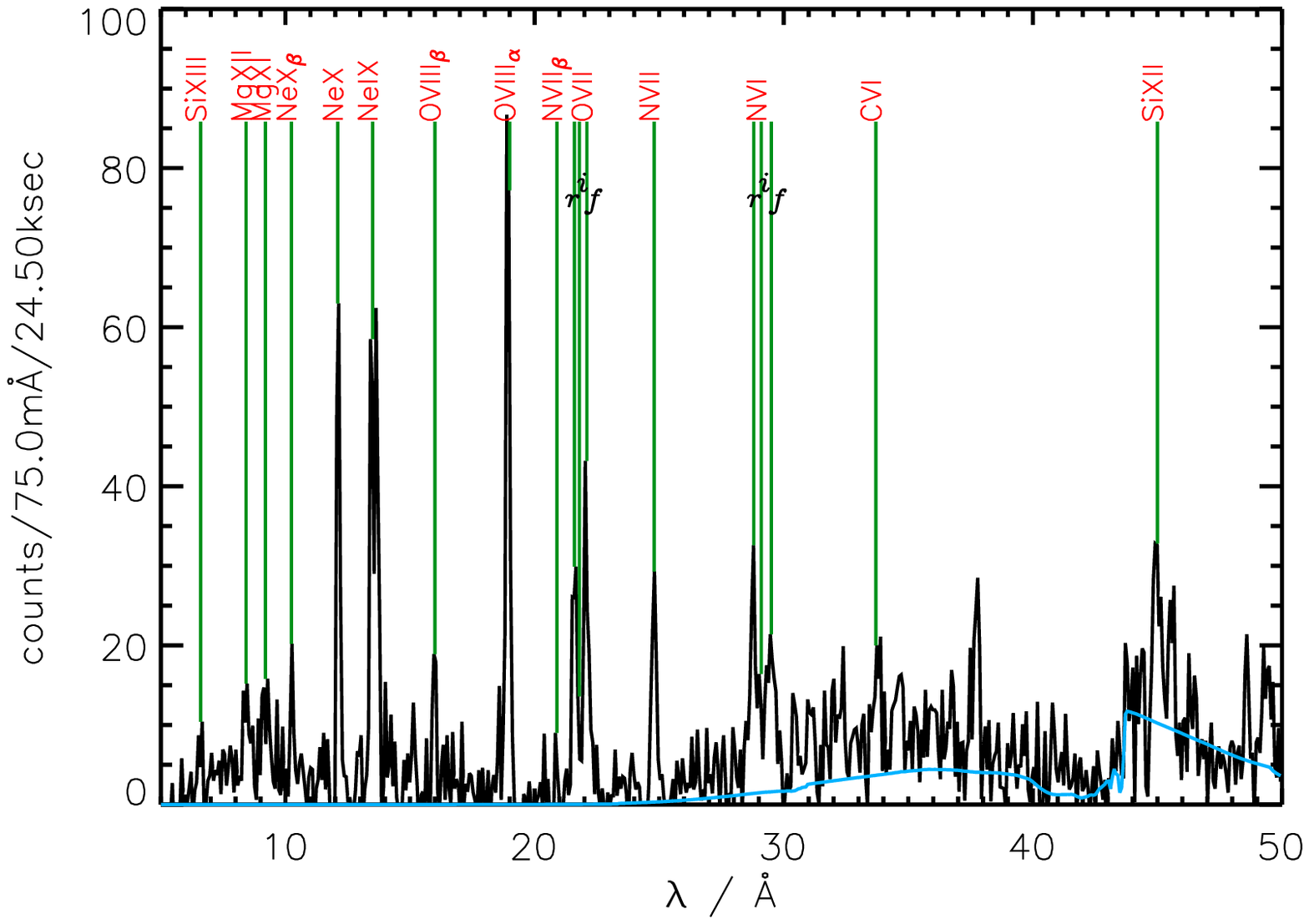}}
\caption{\label{spec}{\it Chandra} LETG spectrum of V382\,Vel (background subtracted).
Only emission lines can be seen with some weak continuum between 30 and 40\,\AA.
A diluted ($R=6000$\,km, $d=2.5$\,kpc) thermal black-body spectrum (absorbed by
$N_{\rm H}=1.2\times 10^{21}$\,cm$^{-2}$) is overplotted on the spectrum and suggests
that the weak continuum is emission from the underlying white dwarf.
The temperature used for the source is $T=3\times 10^5$\,K implying a luminosity of
$2\times 10^{36}$\,erg\,s$^{-1}$.}
\end{figure*}

The outburst of the Classical Nova V382\,Vel was discovered on
1999 May 22 \citep{gil99}. V382\,Vel reached a $V_{\rm max}$ brighter than 3,
making it the brightest nova since V1500\,Cyg in 1975. \cite{vall02}
described optical observations of V382\,Vel and classified the
nova as fast and belonging to the Fe\,{\sc ii} {\it broad} spectroscopic
class. Its distance was estimated to be $1.7\,\pm\,0.34$\,kpc.
Infrared observations detected the [Ne\,{\sc ii}]
emission line at 12.8\,$\mu$m characteristic of the ``neon nova" group and,
subsequently, V382\,Vel was recognized as an ONeMg nova \citep{woodw99}.
An extensive study of the ultraviolet (UV) spectrum was presented by
\cite{shore03}, who analysed spectra obtained with the Space Telescope Imaging
Spectrograph (STIS) on the {\it Hubble Space Telescope} ({\it HST}) in 1999 August as
well as spectra from the {\it Far Ultraviolet Spectroscopic Explorer} ({\it FUSE}).
They found a remarkable resemblance of Nova V382\,Vel
to V1974\,Cyg (1992), thus imposing important constraints on any model for the
nova phenomenon.\\
Following an ``iron curtain" phase (where a significant fraction
of the ultraviolet light is absorbed by elements with atomic numbers around 26)
V382\,Vel proceeded through a stage when P\,Cygni line profiles were detected for
all important UV resonance
lines \citep{shore03}. The line profiles displayed considerable
sub-structure, indicative of fragmentation of the ejecta at the earliest stages of
the outburst. The fits to the UV spectra suggested a distance of
2.5\,kpc, higher than the value given by \cite{vall02}.
From the spectral models, \cite{shore03} were
able to determine element abundances for He, C, N, O, Ne, Mg, Al, and Si,
relative to solar values. While C and Si were found to be
under-abundant, O, Ne, and Al were significantly over-abundant compared to solar
values. From the H Ly$_\alpha$ line at 1216\,\AA, \cite{shore03} estimated
a value for the
hydrogen column-density of $N_{\rm H}=1.2\times 10^{21}$\,cm$^{-2}$.

\subsection{Previous X-ray observations of V382\,Vel}

Early X-ray observations of V382\,Vel were carried out with the {\it RXTE} on day 5.7 by
\cite{ms99}, who did not detect any significant X-ray flux (see Table~\ref{pobs}).
X-rays were first detected from this nova by {\it BeppoSAX} on day 15
\citep[cf., ][]{orio01a} in a very broad band from 0.1--300\,keV. A hard spectrum
was found between 2 and 10\,keV which these authors attributed to emission from
shocked nebular ejecta at a plasma temperature $kT_e\,\sim\,6$\,keV. No soft component
was present in the spectrum.
On day 20.5 \cite{mukai01} found, from {\it ASCA} observations, a highly
absorbed ($N_{\rm H} \sim$ 10$^{23}$\,cm$^{-2}$) bremsstrahlung spectrum with
a temperature $kT_e\sim 10$\,keV. In subsequent observations with {\it RXTE}
(days 31, 35, 50, and 59) the spectrum softened because of decreasing
temperatures ($kT_e\sim 4.0$\,keV to $kT_e\sim 2.4$\,keV on day 59) and
diminishing $N_{\rm H}$ ($7.7\times 10^{22}$\,cm$^{-2}$
on day 31 to $1.7\times 10^{22}$\,cm$^{-2}$ on day 59). \cite{mukai01} argued that the
X-ray emission arose from shocks internal to the nova ejecta.
Like \cite{orio01a}, they did not find any soft component. Six months later,
on 1999 November 23, \cite{orio02} obtained a second {\it BeppoSAX} observation and
detected both a hard component and an extremely bright Super-Soft component. By
that time the (absorption-corrected) flux of the hard component (0.8--2.4\,keV) had
decreased by about a factor of 40, however, the flux below $\sim 2$\,keV
increased by a larger factor \citep{orio02}.

{\it Chandra} Target of Opportunity (ToO) observations were reported by \cite{burw02}.
These authors used both ACIS and the LETG to observe V382\,Vel four times and they gave
an initial analysis of the data obtained. The first ACIS-I observation was obtained on
1999 December 30 and showed that the nova was still in the Super Soft phase
\citep[as seen by ][ about 2 months before]{orio02} and was bright.
This observation was followed on 2000 February 14 by the first
high-resolution X-ray observation of any nova in outburst, using
the Low Energy Transmission Grating (LETG+HRC-S) on
board {\it Chandra}. Its spectral resolution of R$\sim$600
surpasses the spectral resolution of the other X-ray detectors
by factors of up to more than two orders of magnitude.\\
We found that the strong component at $\sim 0.5$\,keV had decreased significantly and
was replaced by a (mostly) emission line spectrum above 0.7keV. The
total flux observed in the 0.4--0.8\,keV range had also declined by a factor of about
100 \citep{burw02}. The grating
observation was followed by two more ACIS-I observations (2000 April 21
and August 14) which showed emission lines and a gradual
fading after the February observation. We therefore conclude that
the hydrogen burning on the surface of the white dwarf in V382\,Vel must have
turned off sometime between 1999 December 30 and 2000 February 14, resulting in a
total duration of 7.5--8 months for the TNR phase. This indicates that the white
dwarf in V382\,Vel has a high mass which is consistent with its
ONeMg nature and short decay time, t$_3$\footnote{The time-scale by which the
visual brightness declines by three orders of magnitude} \citep{krautt96}.
We speculate that hydrogen burning turned off shortly
after the 1999 December 30 observation, since a cooling time
of less than 6 weeks seems extremely short \citep{krautt96}.\\
In this paper we analyse the {\it Chandra} LETG spectrum obtained on 2000 February 14
and present a detailed description of the emission line spectrum. We
describe our measurements and analysis methods in Section~\ref{anal}.
Our results are given in Section~\ref{results}, where we discuss interstellar
hydrogen absorption, line profiles, the lack of iron lines and
line identifications. Our conclusions are summarized in Sections~\ref{disc}
and \ref{conc}.

\section{Analysis}
\label{anal}

The V382\,Velorum observation was carried out on 2000 February
14, 06:23:10\,UT with an exposure time of 24.5\,ksec.
We extracted the {\it Chandra} LETG data from the {\it Chandra} archive and analysed
the preprocessed pha2 file. We calculated effective areas with
the $fullgarf$ task provided by the CIAO software. The background-subtracted
spectrum is shown in Fig.~\ref{spec}. It shows emission lines but no strong
continuum emission. For the measurement of line fluxes we use the original,
non-background-subtracted spectrum. The instrumental background is instead added
to a model spectrum constructed from the spectral model parameters -- wavelengths, line
widths, and line counts. The spectral model consists of the sum of normalized
Gaussian line profiles, one for each emission line, which are each multiplied by
the parameter representing the line counts. Adding the instrumental background to
the spectral model is equivalent to subtracting the background from the original
spectrum (with the assumption that the background is the same under the
source as it was measured adjacent to the source), but is necessary in order to
apply the Maximum Likelihood method conserving Poisson statistics as required
by \cite{cash79} and implemented in {\sc Cora} \citep{newi02}. We also extracted
nine LETG spectra of Capella, a coronal source with strong emission
lines, which are purely instrumentally broadened \citep[e.g.,][]{ness_cap,nebr}.
The combined Capella spectrum is used as a reference in order to detect line
shifts and anomalies in line widths in the spectrum of V382\,Vel. Previous
analyses of Capella \citep[e.g., ][]{argi03} have shown that the lines observed
with {\it Chandra} are at their rest wavelengths.

For the measurement of line fluxes and line widths we use
the {\sc Cora} program \citep{newi02}. This program is a maximum likelihood estimator
providing a statistically correct method to
treat low-count spectra with less than 15 counts per bin. The normal
$\chi^2$ fitting approach requires the spectrum to be rebinned in advance of
the analysis in order to contain at least 15 counts in each spectral bin, thus
sacrificing spectral resolution information. Since background subtraction results
in non-Poissonian statistics, the model spectrum, $c_i$, consists of the sum of
the
background spectrum, $BG$ (instrumental background plus an {\it a priori} given constant
source continuum), and $N_L$ spectral lines, represented by a profile function
$g_{i,j}(\lambda,\sigma)$. Then
\begin{equation}
c_i=BG+\sum^{N_L}_{j=0}A_j\cdot g_{i,j}(\lambda,\sigma)
\end{equation}
with $A_j$ the number of counts in the $j$-th line. The formalism of the
{\sc Cora} program is based on minimizing the (negative) likelihood function
\begin{equation}
\label{like}
{\cal L}= -2 \ \ln P =-2\sum_{i=1}^{N}(-c_i+n_i\ln c_i)
\end{equation}
with $n_i$ being the measured (non-subtracted)
count spectrum, and $N$ the number of spectral bins.
We model the emission lines as Gaussians representing only instrumentally
broadened emission lines in the case of the coronal source Capella
\citep[e.g.,][]{ness_cap}. In V382\,Vel the lines are Doppler broadened, but,
as will be discussed in Section~\ref{lprop}, we have no reason to use more refined
profiles to fit the emission lines in our spectrum. The {\sc Cora}
program fits individual lines under the assumption of a locally constant
underlying source continuum. For our purposes we set this continuum
value to zero for all fits, and account for the continuum emission between
30 and 40\,\AA\ by adding the model flux to the instrumental background.
All line fluxes are then measured above the continuum value at the respective
wavelength (see Fig.~\ref{spec}).

\section{Results}
\label{results}
\subsection{Interstellar absorption}
\label{nh}

\begin{figure}
  \resizebox{\hsize}{!}{\includegraphics{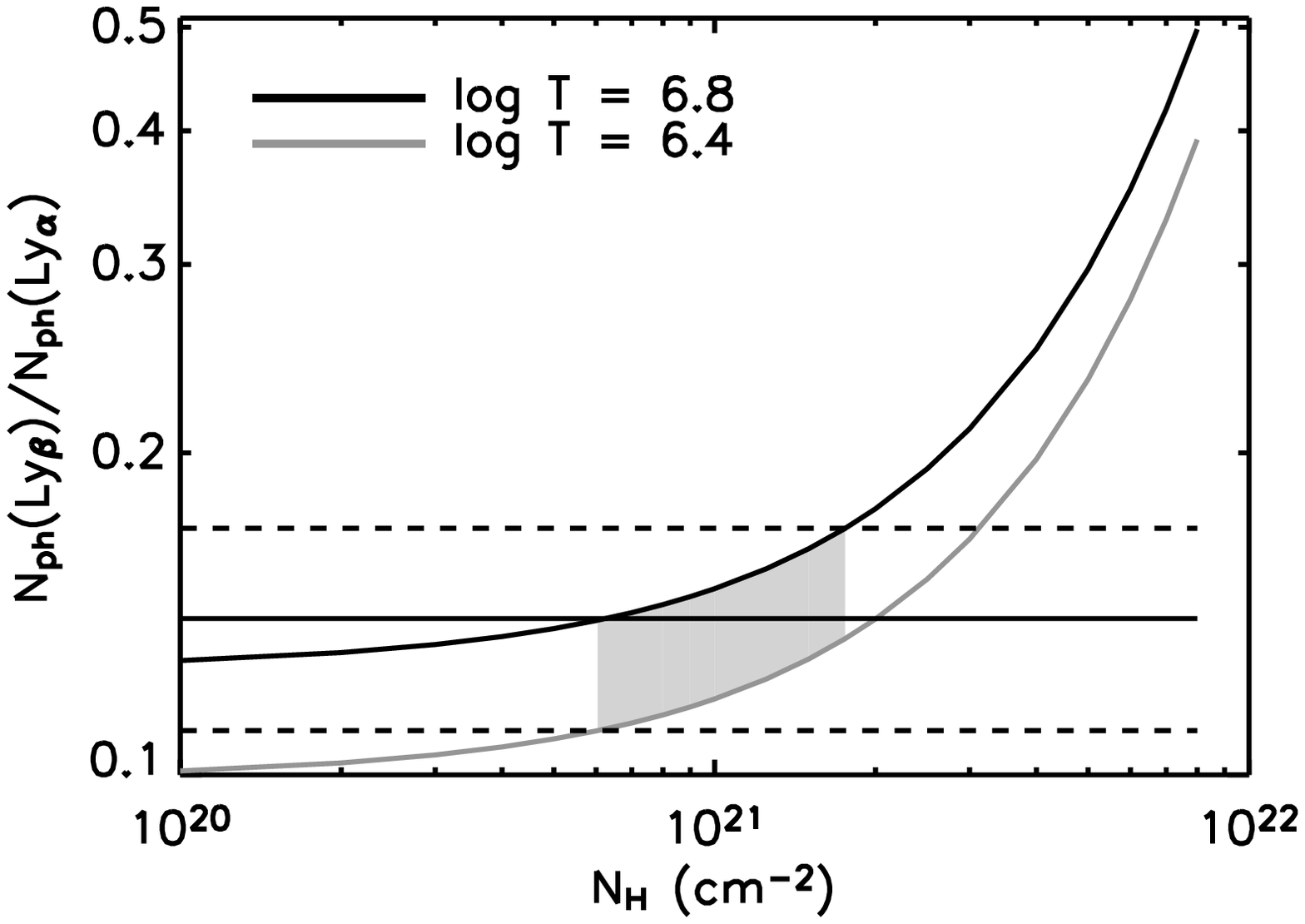}}
\caption{\label{nh_abs}Dependence of the photon flux ratio for O\,{\sc viii}
Ly$_\beta$/Ly$_\alpha$ on $N_{\rm H}$. As $N_{\rm H}$ increases, Ly$_\alpha$ is
more absorbed than Ly$_\beta$. The grey shaded area marks the range of $N_{\rm H}$
consistent with our measurement of O\,{\sc viii} Ly$_\beta$/Ly$_\alpha$, with
associated errors.}
\end{figure}

\begin{figure*}
 \resizebox{\hsize}{!}{\includegraphics{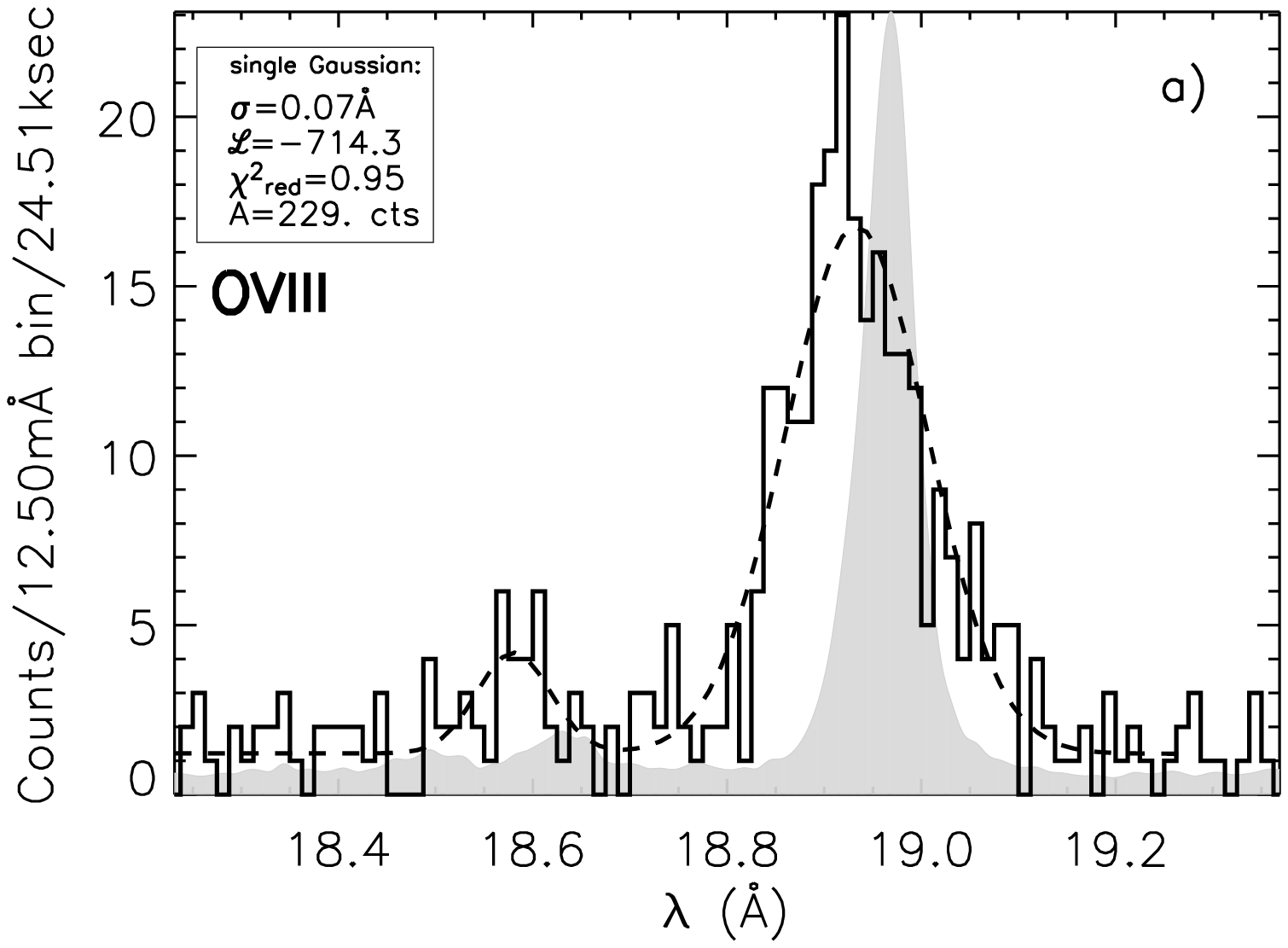}\includegraphics{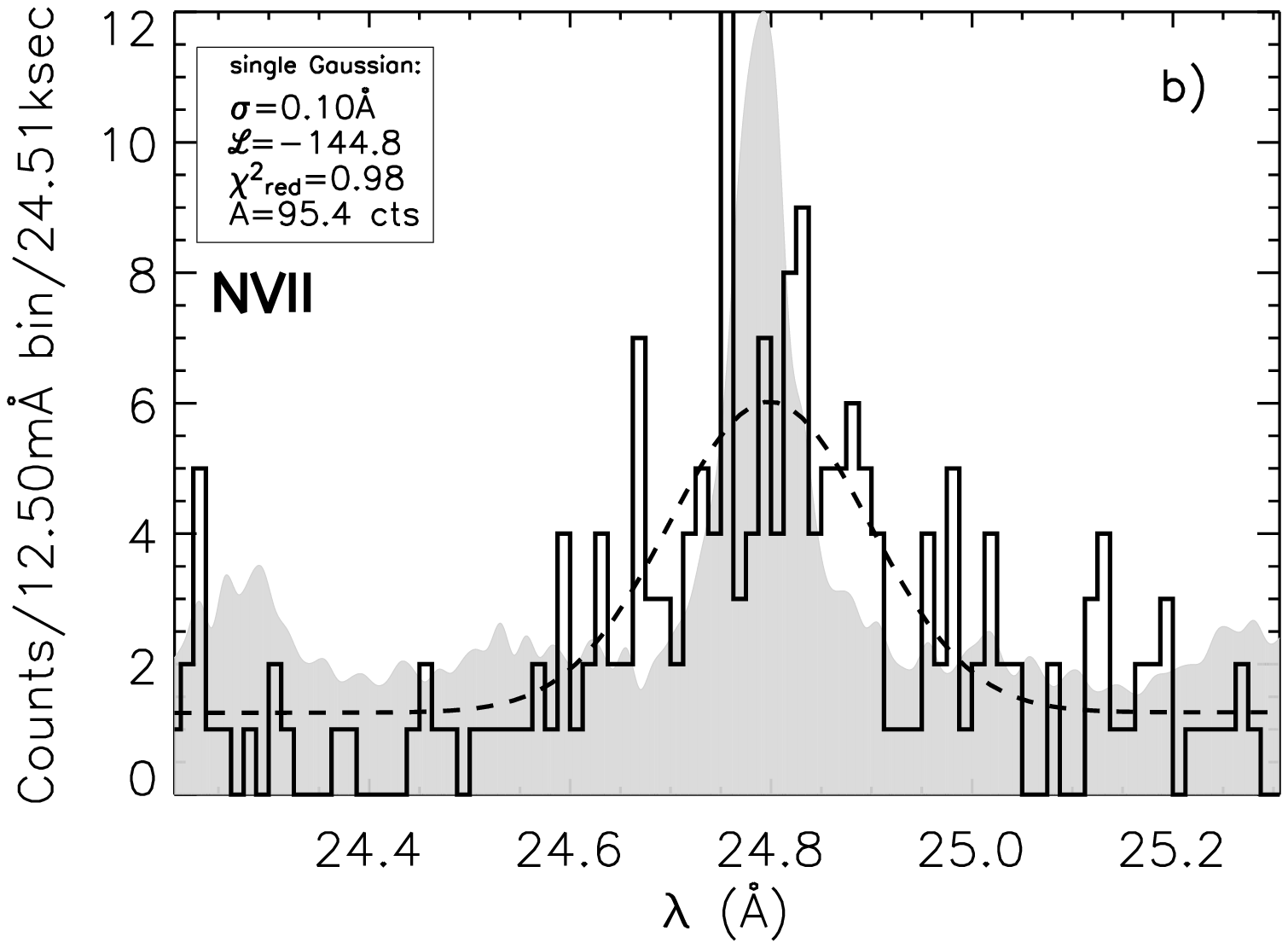}}
\vspace{-.9cm}

 \resizebox{\hsize}{!}{\includegraphics{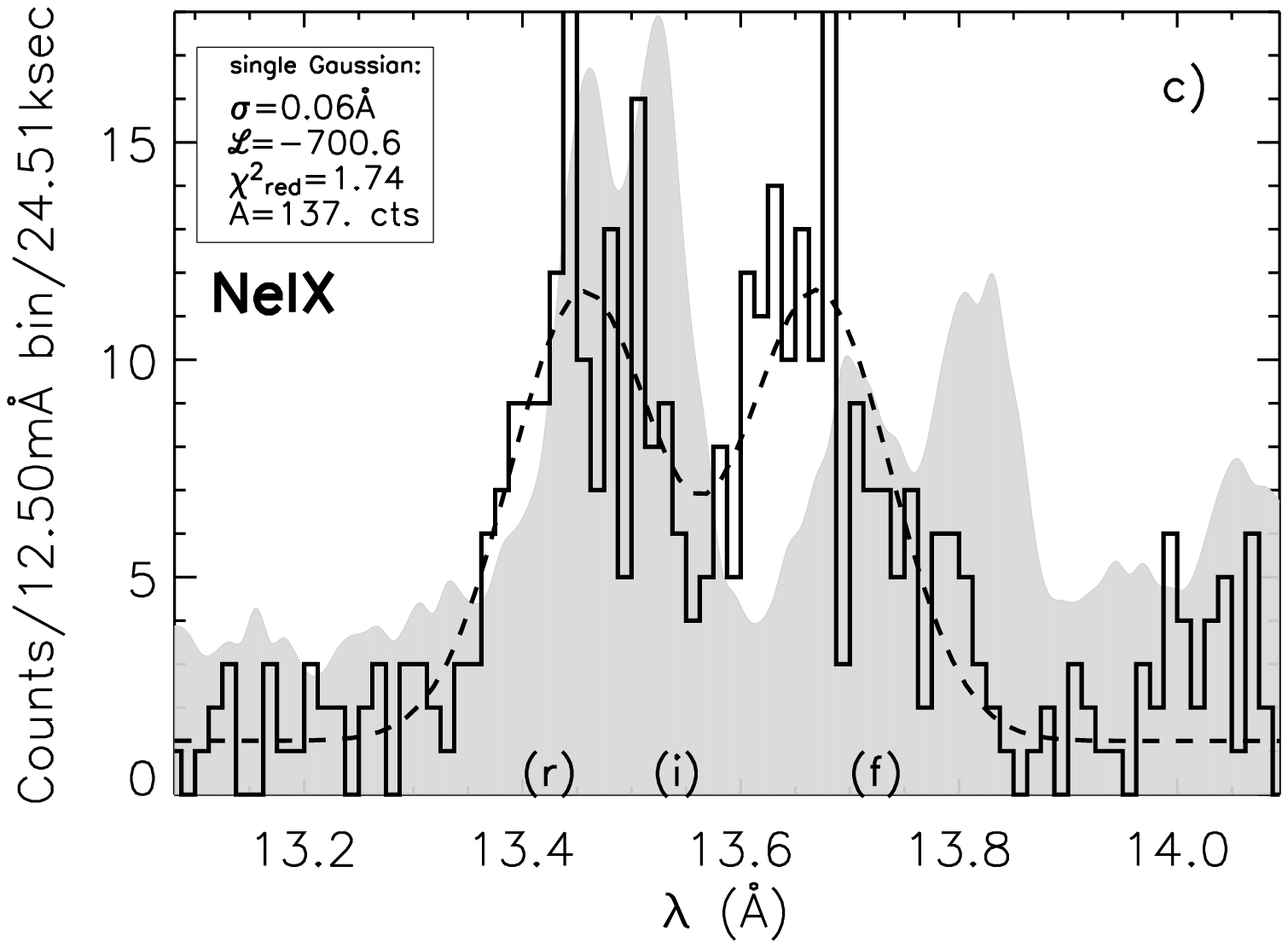}\includegraphics{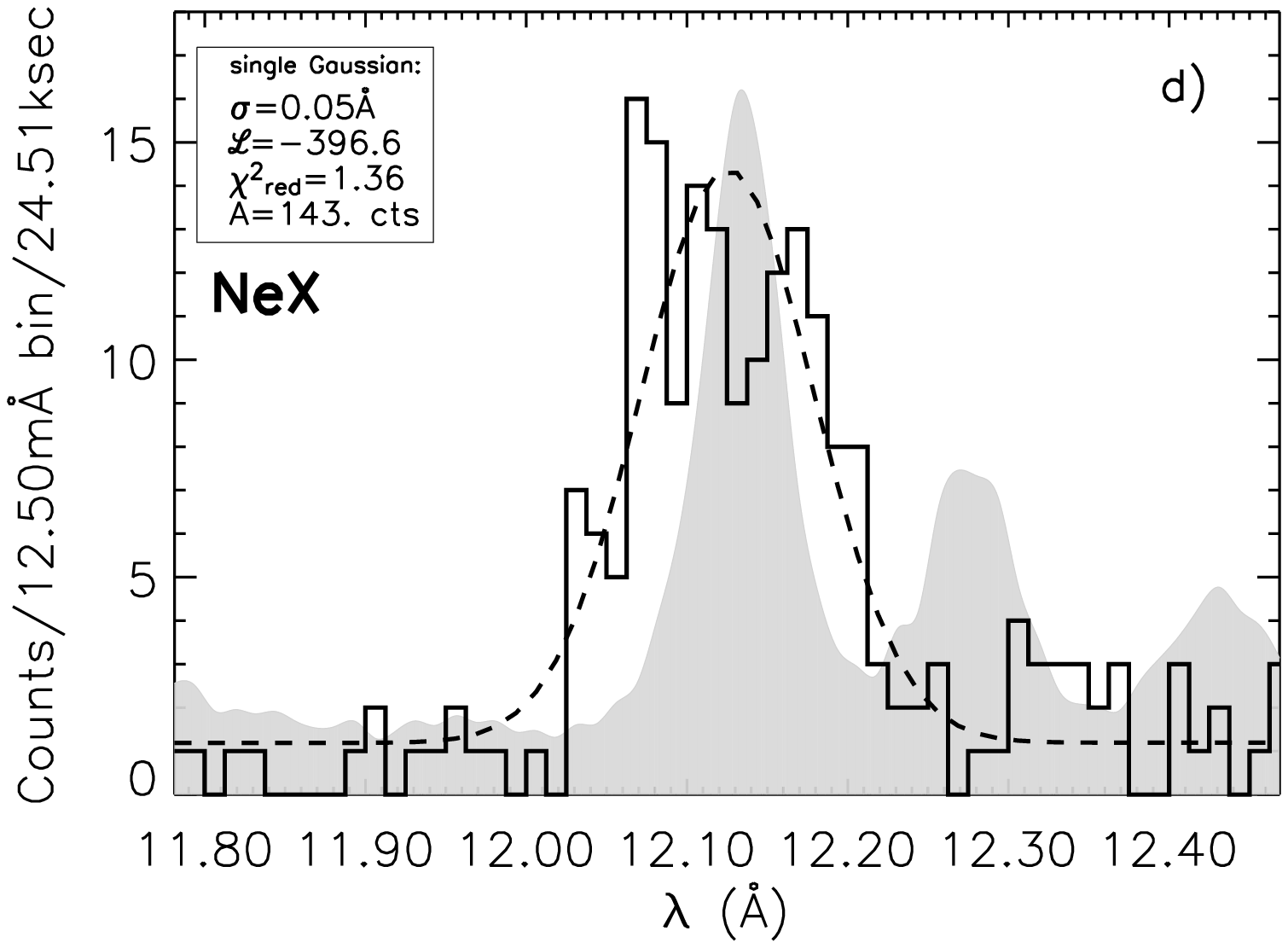}}
\caption[]{\label{lines}Profile analysis of the strongest, isolated lines. Best-fitting
profiles using a single, broadened Gaussian line template are also shown. The light curves
are arbitrarily scaled smoothed LETG spectra of Capella representing the instrumental
resolving power. The legend gives the fit parameters, where $sigma$ is the line width,
$\ell$ is the best likelihood value, $\chi^2$ a goodness parameter and A the number
of counts in the lines.}
\end{figure*}

\begin{figure*}
 \resizebox{\hsize}{!}{\includegraphics{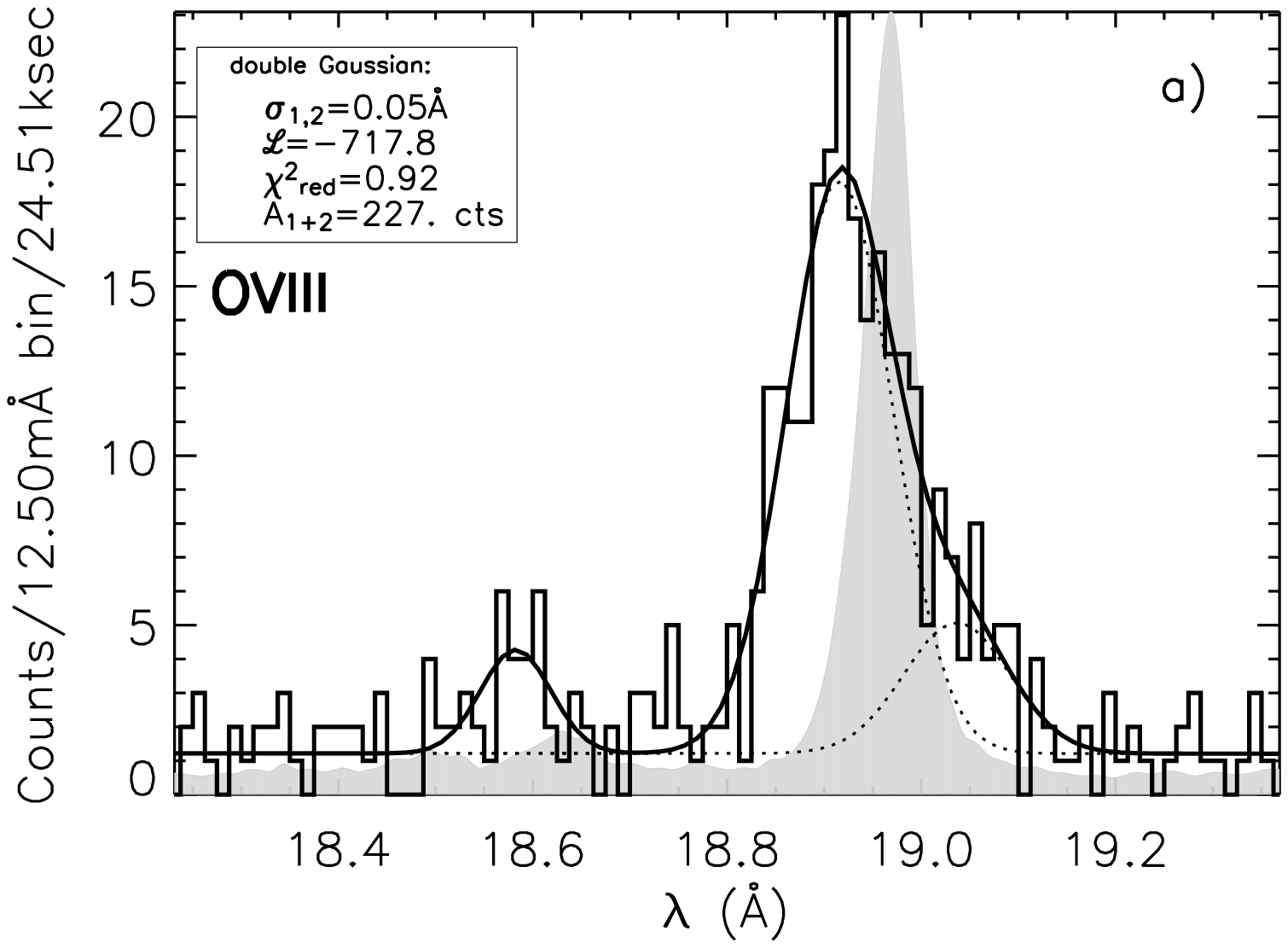}\includegraphics{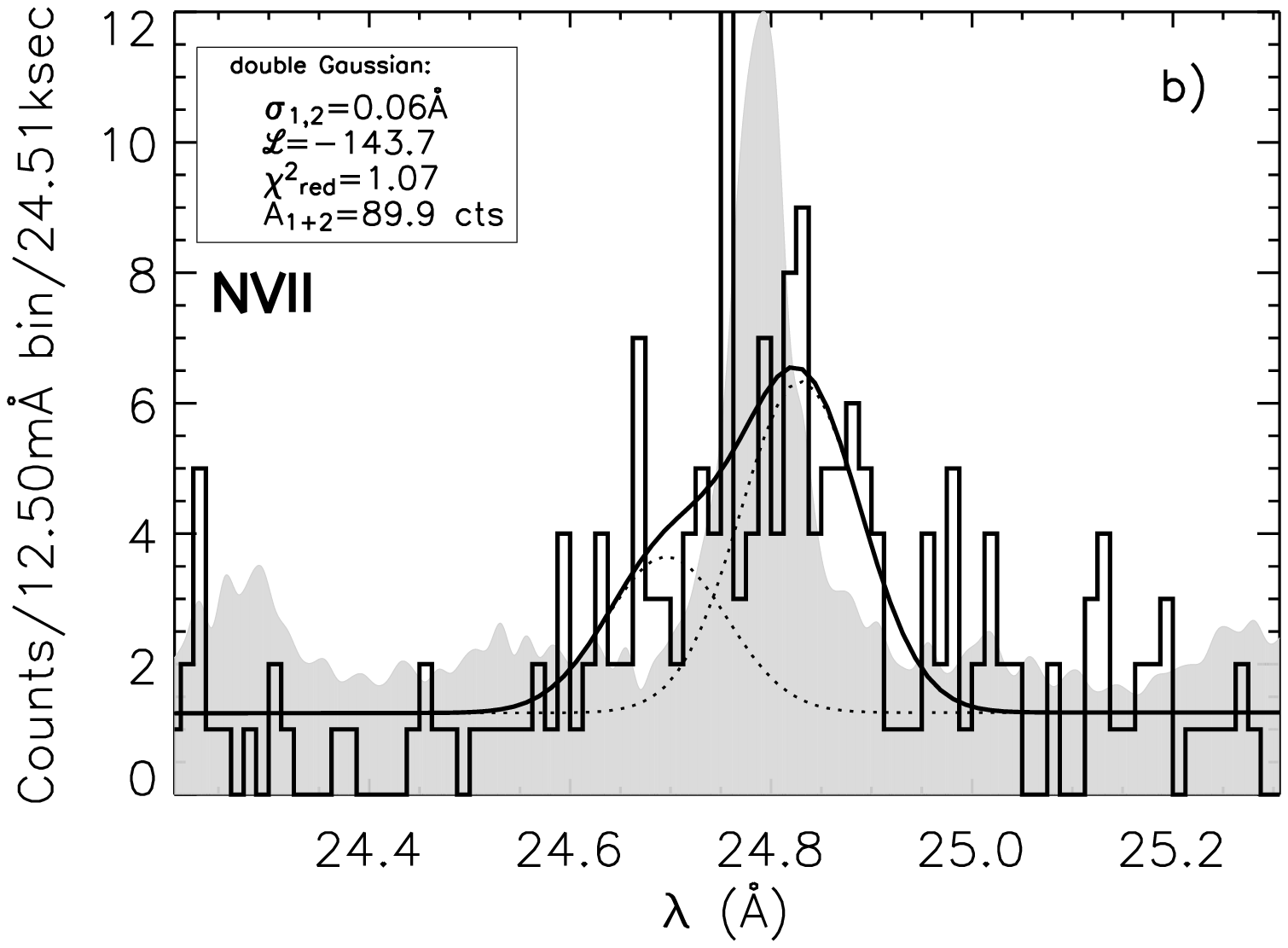}}
\vspace{-.9cm}

 \resizebox{\hsize}{!}{\includegraphics{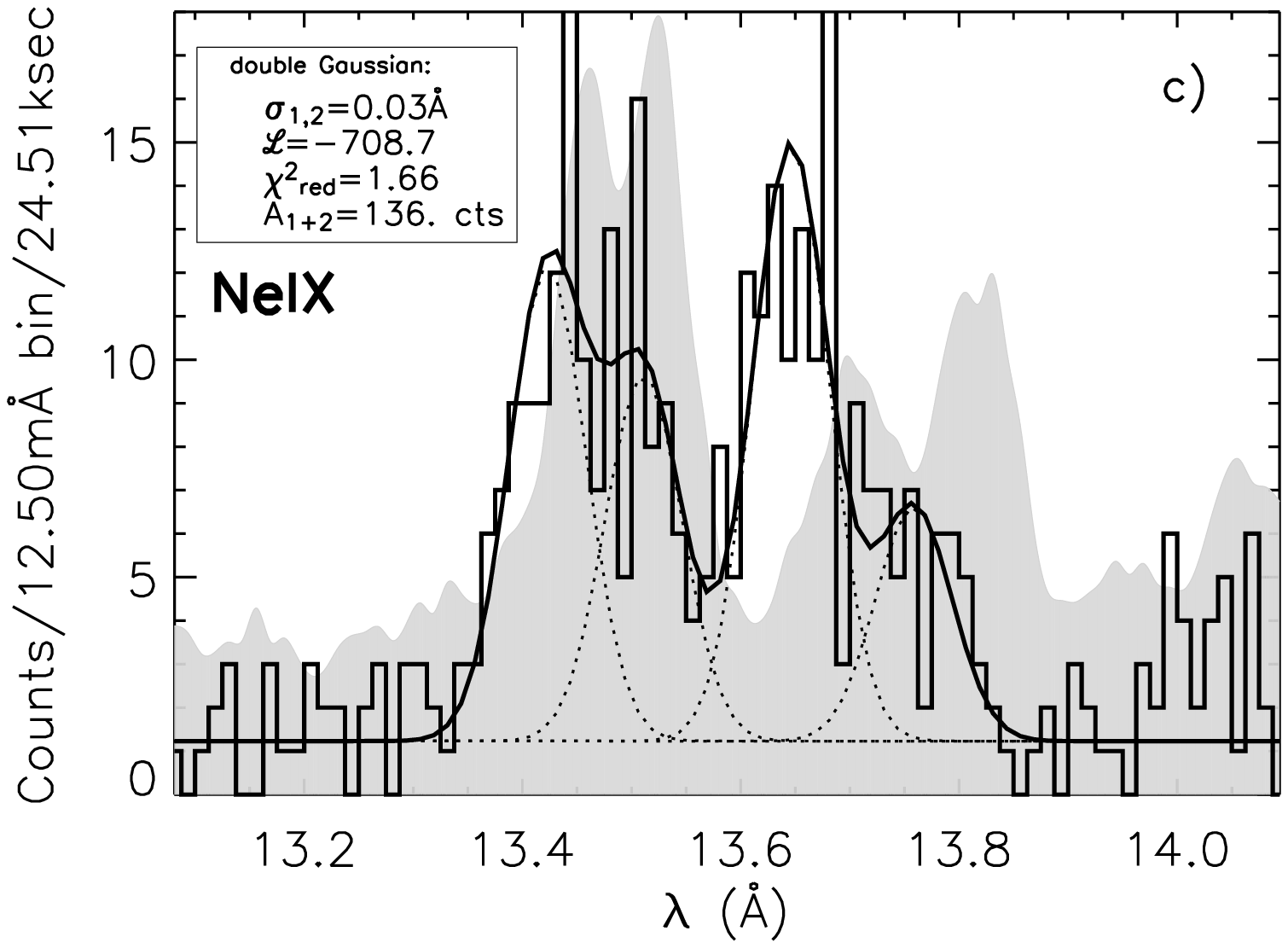}\includegraphics{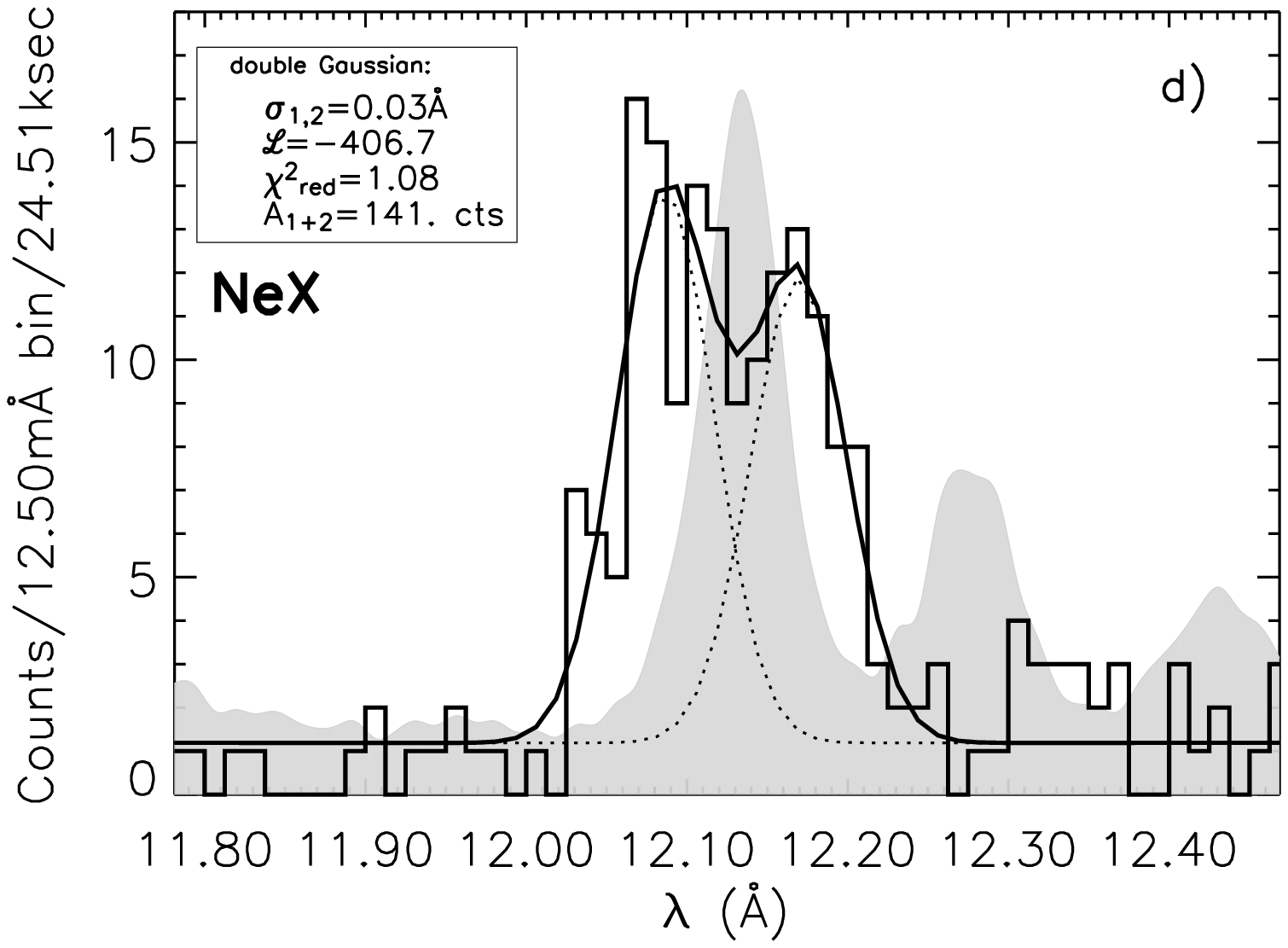}}
\caption[]{\label{lines_doub}Profile analysis of the strongest, isolated lines. The
best-fitting profiles
using two instrumentally broadened Gaussian lines (dotted lines mark the individual
components) can be compared with Fig.~\ref{lines}. The fit parameters are defined in
the caption of Fig.~\ref{lines}.}
\end{figure*}

Since V382\,Vel is located at a considerable distance along the Galactic plane,
substantial interstellar absorption is expected to affect its soft X-ray spectra.
In the early phases, spectral fits to {\it RXTE} and {\it ASCA} data taken immediately
after the outburst (1999 May 26 to July 18) showed a decrease of
$N_{\rm H}$ from $10^{23}$\,cm$^{-2}$ to $<5.1\times
10^{22}$\,cm$^{-2}$ \citep{mukai01}. {\it BeppoSAX} observations
carried out in 1999 November revealed an $N_{\rm H}$ around
$2\times 10^{21}$\,cm$^{-2}$ \citep{orio02}. In 1999 August, the expanding
shell was already transparent to UV emission and \cite{shore03} measured an
$N_{\rm H}$ of $1.2\times 10^{21}$\,cm$^{-2}$.

The high-resolution X-ray spectrum offers a new opportunity to determine
$N_{\rm H}$ from the observed ratio of the O\,{\sc viii} Ly$_{\alpha}$ and
Ly$_{\beta}$ line fluxes. This line ratio depends only weakly on temperature,
but quite sensitively on $N_{\rm H}$ (see Fig.~\ref{nh_abs}). From the photon flux
ratio of 0.14\,$\pm$\,0.03 (which has been corrected for the effective area), we
infer an equivalent $N_{\rm H}$-column-density between
$6\times 10^{20}$\,cm$^{-2}$ (assuming $\log T=6.4$) and $2\times 10^{21}$\,cm$^{-2}$
(assuming $\log T=6.8$), which is consistent with the value determined by
\cite{shore03}. This value appears to represent the true interstellar absorption,
rather than absorption intrinsic to the nova itself.
For our spectral analyses we adopt this value and calculate transmission coefficients
from photoelectric absorption cross-sections using the polynomial fit coefficients
determined by \cite{bamabs}. We assume standard cosmic abundances from \cite{agrev89}
as implemented in the software package PINTofALE \citep{pintofale}.

\subsection{Continuum emission}

All previous X-ray spectra of V382\,Vel were probably dominated by continuum emission
although we cannot rule out the possibility that the continuum consisted of a large
number of overlapping emission lines. In contrast, the LETG spectrum from 2000 Feb 14
does not exhibit continuum emission over the entire wavelength range.
However, some 'bumps' can be seen in Fig.~\ref{spec} at around 35\,\AA,
which could be interpreted as weak continuum emission, or could be the
result of a large number of weak overlapping lines. The low count rate around 44\,\AA\
is due to C absorption in the UV shield, filter and detector coating at this
wavelength. We calculated a diluted thermal black-body spectrum for the WD,
absorbed with $N_{\rm H}=1.2\times 10^{21}$\,cm$^{-2}$ (Section~\ref{nh}). The
continuum is calculated assuming a WD radius of 6000\,km, a distance of 2.5\,kpc
and a temperature of $3\times 10^5$\,K. Given the possible presence of weak lines,
this temperature is an upper limit. The intrinsic (integrated) luminosity implied by
the black-body source is $L_{\rm BB}=2\times 10^{36}$\,erg\,s$^{-1}$ (which corresponds
to an X-ray luminosity of $\sim 10^{28}$\,erg\,s$^{-1}$. The black-body spectrum was
then multiplied
by the exposure time and the effective areas in order to compare it with the (rebinned)
count spectrum shown in Fig.~\ref{spec}. We do not use the parameters of the
black-body source in our further analysis, but it is clear that even the highest possible
level of continuum emission is not strong enough to excite lines by photoexcitation.
Also, no high-energy photons are observed that could provide any significant
photoexcitation or photoionization. In the further analysis, we therefore assume that
the lines are exclusively produced by collisional ionization and excitation.
We point out that, given the uncertainty in the assumed radius and distance, our
upper limit to the black-body temperature is still consistent with the lower limit
of $2.3\times 10^5$\,K found by \cite{orio02} by
fitting WD NLTE atmospheric models to the spectrum obtained in the second
{\it BeppoSAX} observation.

\subsection{Analysis of line properties}
\label{lprop}

In Fig.~\ref{lines}, we show the spectral ranges around the strong resonance
lines of the H-like ions O\,{\sc viii} at 18.97\,\AA, N\,{\sc vii} at
24.78\,\AA, and Ne\,{\sc x} at 12.13\,\AA, the resonance line of the
He-like ion Ne\,{\sc ix} at 13.45\,\AA\ and the Ne\,{\sc ix} forbidden line
at 13.7\,\AA. In order to illustrate the instrumental line profile, we plot
an LETG spectrum of Capella in a lighter shade (the
Capella spectrum is arbitrarily scaled to give overall intensities that can
be plotted on the same scales as those for Nova V382\,Vel).
Clearly, the lines in V382\,Vel are significantly broadened compared to the
emission lines in the Capella spectrum. We test two
hypotheses to describe the broadening of the lines. First, we use a
single Gaussian with an adjustable Gaussian line width parameter
$\sigma$ (defined through $I=I_0e^{-0.5(\lambda-\lambda_0)^2/\sigma^2}$). Secondly,
we use a double Gaussian line profile with adjustable wavelengths.
We show the best-fitting curves of single profiles in Fig.~\ref{lines} and those
of double profiles in Fig.~\ref{lines_doub} (individual components are dotted).
For O\,{\sc viii}, we also fit the N\,{\sc vii} Ly$_\beta$ line at 18.6\,\AA\ with
a single Gaussian. In the legends we provide the goodness parameters ${\cal L}$ from
equation~\ref{like} and $\chi^2$, which is only given for information,
because it gives a quantitative figure of the quality of agreement, while the likelihood
value ${\cal L}$ is only a relative number. To obtain the best fit we minimized
${\cal L}$, since our spectrum contains fewer than 15 counts in
most bins, and $\chi^2$ does not qualify as a goodness criterion \citep{cash79}.
The fit parameters, line width, $\sigma$, and line counts, $A$, are also given.
For all fits, those models with two Gaussian lines return slightly better
likelihood values. For O\,{\sc viii} and
N\,{\sc vii} each component of the double Gaussian is broader than the
instrumental line width indicating that the O\,{\sc viii}
and N\,{\sc vii} emission originates from at least three different emission regions
supporting the fragmentation scenario suggested by \cite{shore03}. The O\,{\sc viii}
line appears blue-shifted with respect to the rest wavelengths with only weak
red-shifted O\,{\sc viii} emission. The N\,{\sc vii} line is split into several
weaker components
around the central line position, indicating the highest degree of fragmentation,
although the noise level is also higher. At longer wavelengths, fragmentation in
velocity space can be better resolved owing to the increasing spectral resolution.
The lines from both Ne\,{\sc x} and Ne\,{\sc ix} seem to be
confined to two distinctive fragmentation regions. For other elements, this exercise
cannot be carried out because the respective lines are too faint. From the Gaussian
line widths of the single-line fits, converted to FWHM, we derive Doppler
velocities and find 2600\,km\,s$^{-1}$, 2800\,km\,s$^{-1}$, 2900\,km\,s$^{-1}$, and 3100\,km\,s$^{-1}$ for
O\,{\sc viii}, N\,{\sc vii}, Ne\,{\sc x}, and Ne\,{\sc ix}, respectively.
These are roughly consistent within the errors ($\sim 200$\,km\,s$^{-1}$) but they are
lower than the expansion velocities reported by \cite{shore03}, who found
$4000$\,km\,s$^{-1}$ (FWHM) from several UV emission lines measured some eight months
earlier. What may be happening is that the density and emissivity of the fastest
moving material is decreasing rapidly so that over time we see through it
to slower, higher-density inner regions.

\begin{figure}
  \resizebox{\hsize}{!}{\includegraphics{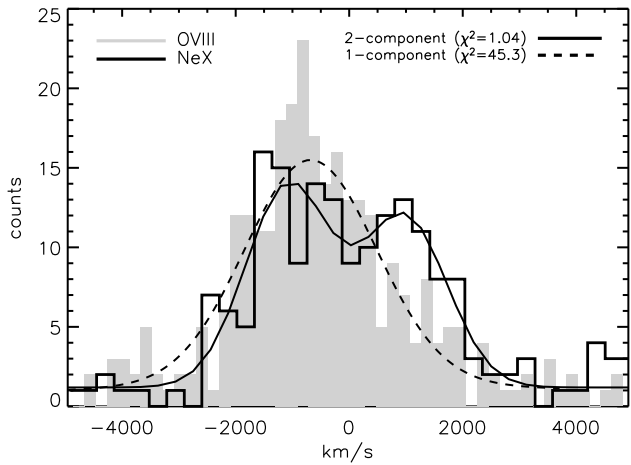}}
\caption{\label{o8ne10}Direct comparison of velocity structure in O\,{\sc viii}
(light shade) and Ne\,{\sc x} (dark line). The best-fitting double-line profile from
Fig.~\ref{lines_doub}c (solid) is compared to the rescaled single-line profile of
Fig.~\ref{lines}a (dashed).}
\end{figure}

Fig.~\ref{o8ne10} shows a comparison of the profiles of the O\,{\sc viii} and
Ne\,{\sc x} lines in velocity space
($v=c(\lambda-\lambda_0)/\lambda_0$) with $\lambda_0=18.97$\,\AA\ for O\,{\sc viii}
and $\lambda_0=12.13$\,\AA\ for Ne\,{\sc x}. The shaded O\,{\sc viii} line
is blue shifted, while Ne\,{\sc x} shows a red-shifted and a blue-shifted component
of more equal intensity at roughly similar velocities. In order to quantitatively
assess the agreement between the two profiles we attempted to adjust the single-line
profile of the O\,{\sc viii} line (Fig.~\ref{lines}a) to the Ne\,{\sc x}
line, but found unsatisfactory agreement. We adjusted only the number of counts but
not the wavelength or line width of the O\,{\sc viii} template. The difference in
$\chi^2$ given in the upper right legend clearly shows that the profiles are different.
This can be due either to different velocity structures in the respective elements or
to different opacities in the lines. In the latter case the red-shifted component of
O\,{\sc viii} would have to be absorbed while the plasma in the line-of-sight
remained transparent to the red-shifted component of Ne\,{\sc x}.

\subsection{Line identifications}
\label{lineid}

\begin{table}
\renewcommand{\arraystretch}{1.1}
\caption{\label{ltab}Line fluxes (not corrected for N$_{\rm H}$) detected in the
LETG spectrum of V382\,Vel with identifications using APEC (exposure time
$\Delta$t=24.5\,ksec)}
\begin{flushleft}
{\scriptsize
\begin{tabular}{p{.4cm}p{.4cm}p{1.4cm}p{.4cm}p{.5cm}p{.1cm}r}
\hline
$\lambda$ & $\sigma^{[a]}$ & line flux$^{[b]}$ & A$_{\rm eff}$ & $\lambda^{[c]}$ & log($T_
M^{[d]}$) & ID \\
(\AA) & (\AA) & & (cm$^2$) & (\AA) & MK & \\
\hline
6.65 &  0.06 & 4.79\,$\pm$\,1.80 &   44.26 &  6.65 &  7.0 &  Si\,{\sc xiii} (He$_r$)\\
 &  &  &  &  6.74 &  7.0 &  Si\,{\sc xiii} (He$_f$)\\
8.45 &  0.10 & 12.0\,$\pm$\,2.50 &   37.98 &  8.42 &  7.0 &  Mg\,{\sc xii} (Ly$_\alpha$)\\
9.09 &  0.04 & 6.68\,$\pm$\,1.85 &  32.27 &  9.17 &  6.8 &  Mg\,{\sc xi} (He$_r$)\\
9.30 &  0.04 & 6.13\,$\pm$\,1.72 &   32.27 &  9.31 &  6.8 &  Mg\,{\sc xi} (He$_f$)\\
10.24 &  0.04 & 8.93\,$\pm$\,1.99 &   28.62 &   10.24 &  6.8 &  Ne\,{\sc x} (Ly$_\beta$)\\
12.12 &  0.05 & 31.4\,$\pm$\,2.98 &   28.68 &   12.13 &  6.8 &  Ne\,{\sc x} (Ly$_\alpha$)\\
13.45 &  0.05 & 24.7\,$\pm$\,2.58 &   29.37 &   13.45 &  6.6 &  Ne\,{\sc ix} (He$_r$)\\
13.66 &  0.07 & 27.6\,$\pm$\,2.72 &   29.42 &   13.70 &  6.6 &  Ne\,{\sc ix} (He$_f$)\\
15.15 &  0.05 & 3.13\,$\pm$\,1.14 &   30.37 &  15.18 &  6.5 &  O\,{\sc viii} (Ly$_\gamma$)\\
15.98 &  0.06 & 7.22\,$\pm$\,1.45 &   29.98 &   16.01 &  6.5 &  O\,{\sc viii} (Ly$_\beta$)\\
18.58 &  0.07 & 4.63\,$\pm$\,1.23 &   26.16 &   18.63 &  6.3 &  O\,{\sc vii} (1$\rightarrow$13)\\
18.93 &  0.07 & 36.0\,$\pm$\,2.60 &   26.61 &   18.97 &  6.5 &  O\,{\sc viii} (Ly$_\alpha$)\\
20.86 &  0.06 & 2.89\,$\pm$\,1.32 &   17.33 &   20.91 &  6.3 &  N\,{\sc vii} (Ly$_\beta$)\\
21.61 &  0.09 & 23.4\,$\pm$\,2.67 &   17.24 &   21.60 &  6.3 &  O\,{\sc vii} (He$_r$)\\
22.07 &  0.11 & 24.1\,$\pm$\,2.68 &   16.98 &   22.10 &  6.3 &  O\,{\sc vii} (He$_f$)\\
24.79 &  0.10 & 17.6\,$\pm$\,2.27 &   16.84 &   24.78 &  6.3 &  N\,{\sc vii} (Ly$_\alpha$)\\
28.77 &  0.11 & 18.6\,$\pm$\,2.33 &   15.33 &   28.79 &  6.2 &  N\,{\sc vi} (He$_r$)\\
29.04 &  0.06 & 4.80\,$\pm$\,1.44 &   15.10 &   29.08 &  6.1 &  N\,{\sc vi} (He$_i$)\\
29.49 &  0.18 & 22.1\,$\pm$\,2.80 &   14.21 &   29.53 &  6.1 &  N\,{\sc vi} (He$_f$)\\
30.44 &  0.06 & 6.80\,$\pm$\,1.77 &   12.02 &  30.45  &  6.5  &  (?) S\,{\sc xiv} (1$\rightarrow$5,6)\\
31.08 &  0.10 & 7.53\,$\pm$\,1.78 &   14.34 &   &   &  ?\\
32.00 &  0.12 & 7.25\,$\pm$\,1.84 &   13.63 &   &   &  ?\\
32.34 &  0.09 & 8.72\,$\pm$\,1.90 &   13.42 &   32.42  &   6.5  &  (?) S\,{\sc xiv} (2$\rightarrow$7)\\
33.77 &  0.13 & 13.9\,$\pm$\,2.33 &   12.74 &   33.73 &  6.1 &  C\,{\sc vi} (Ly$_\alpha$)\\
34.62 &  0.15 & 12.7\,$\pm$\,2.36 &   12.58 &    &   &  ?\\
37.65 &  0.15 & 28.7\,$\pm$\,3.01 &   10.30 &    &   &  ?\\
44.33 &  0.12 & 3.38\,$\pm$\,0.81 &   26.03 &   &   &  ?\\
44.97 &  0.16 & 9.17\,$\pm$\,1.10 &   25.76 &   &   &  ?\\
45.55 &  0.14 & 5.50\,$\pm$\,0.93 &   25.65 &   45.52 &  6.3 &  (?) Si\,{\sc xii} (2$\rightarrow$4)\\
 &  &  &  &   45.69 &  6.3 &  (?) Si\,{\sc xii} (3$\rightarrow$4)\\
48.59 &  0.09 & 3.53\,$\pm$\,0.68 &   24.21 &    &   &  ?\\
49.41 &  0.10 & 4.31\,$\pm$\,0.74 &   24.52 &    &   &  ?\\
\hline
\end{tabular}
\\
$^{[a]}$Gaussian width parameter; FWHM=$2\sigma \sqrt{2\ln 2}.$
$^{[b]}10^{-14}$\,erg\,cm$^{-2}$\,s$^{-1}$.\\
$^{[c]}$Theor. wavelength from APEC.\hspace{.2cm}
$^{[d]}$Optimum formation temperature.
}
\renewcommand{\arraystretch}{1}
\end{flushleft}
\end{table}

After measuring the properties of the strongest identified emission lines,
we scanned the complete spectral
range for all emission lines. We used a modified spectrum, where the continuum
spectrum shown in Fig.~\ref{spec} is added to the instrumental background such
that, with our measurement method provided by {\sc Cora}, we measure counts above
this continuum background value. In Table~\ref{ltab} we list all lines detected
with a significance level higher than $4\sigma$ (99.9 per cent). We list the central
wavelength, the Gaussian line width and the flux contained in each
line (not corrected for N$_{\rm H}$). We also list the sum of effective areas from
both dispersion directions at
the observed wavelength, obtained from the {\sc Ciao} tool {\sc fullgarf}
which can be used to recover count rates. We used the Atomic Plasma Emission Code
(APEC)\footnote{Version 2.0; available at http://cxc.harvard.edu/atomdb}
and extracted all known lines within two line
widths of each observed line. Usually, more than one line candidate was
found. The various possibilities were ranked according to their
proximity to the observed wavelength and emissivities in the APEC database
(calculated at very low densities). A good fit to the observed wavelength is the
most important factor, taking into account the blue-shifts of the well identified
lines. Theoretical line fluxes can be predicted depending on the emission measure
distribution and the element abundances and are therefore less secure.

\begin{figure*}
  \resizebox{\hsize}{!}{\includegraphics{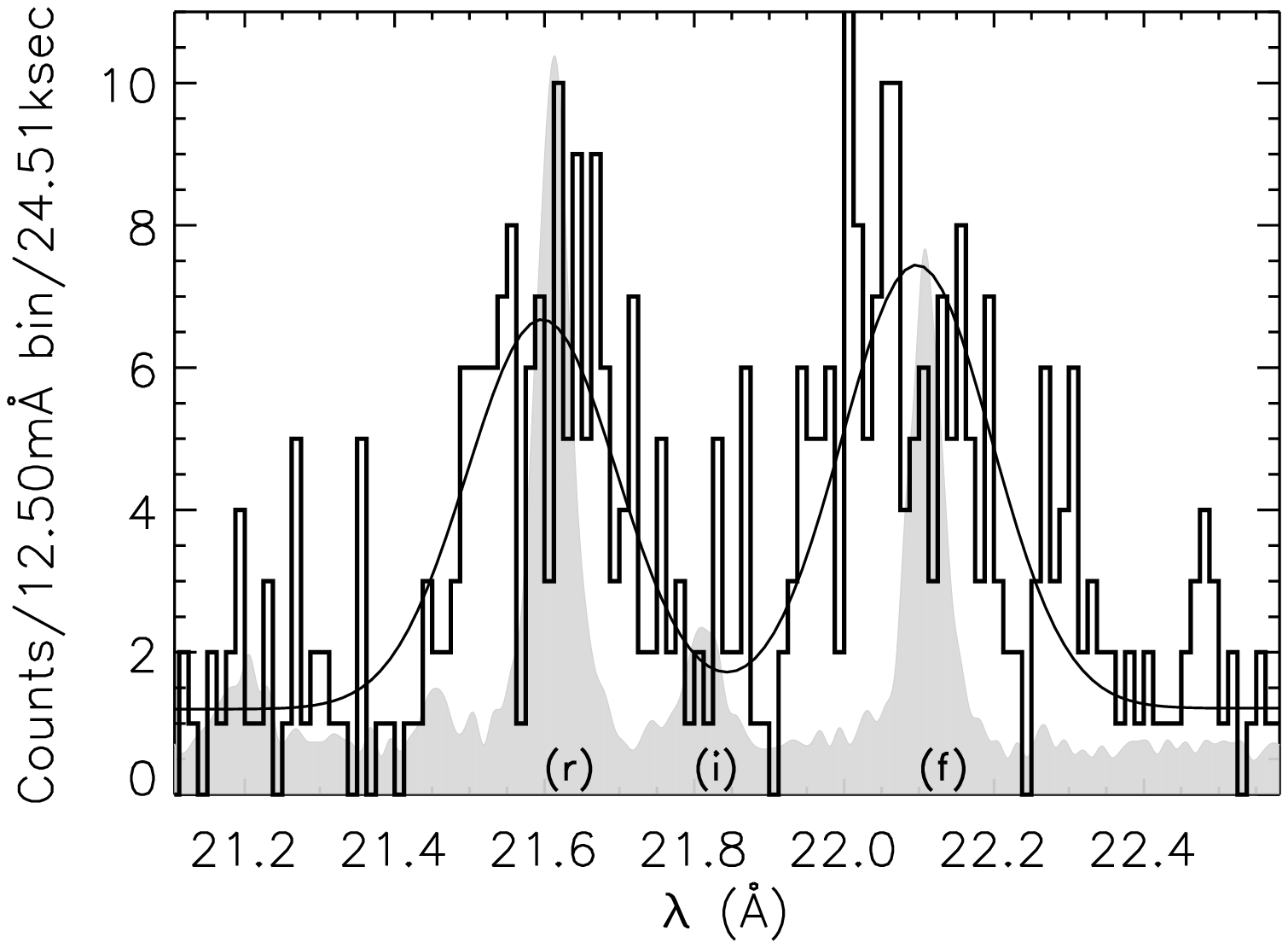}\includegraphics{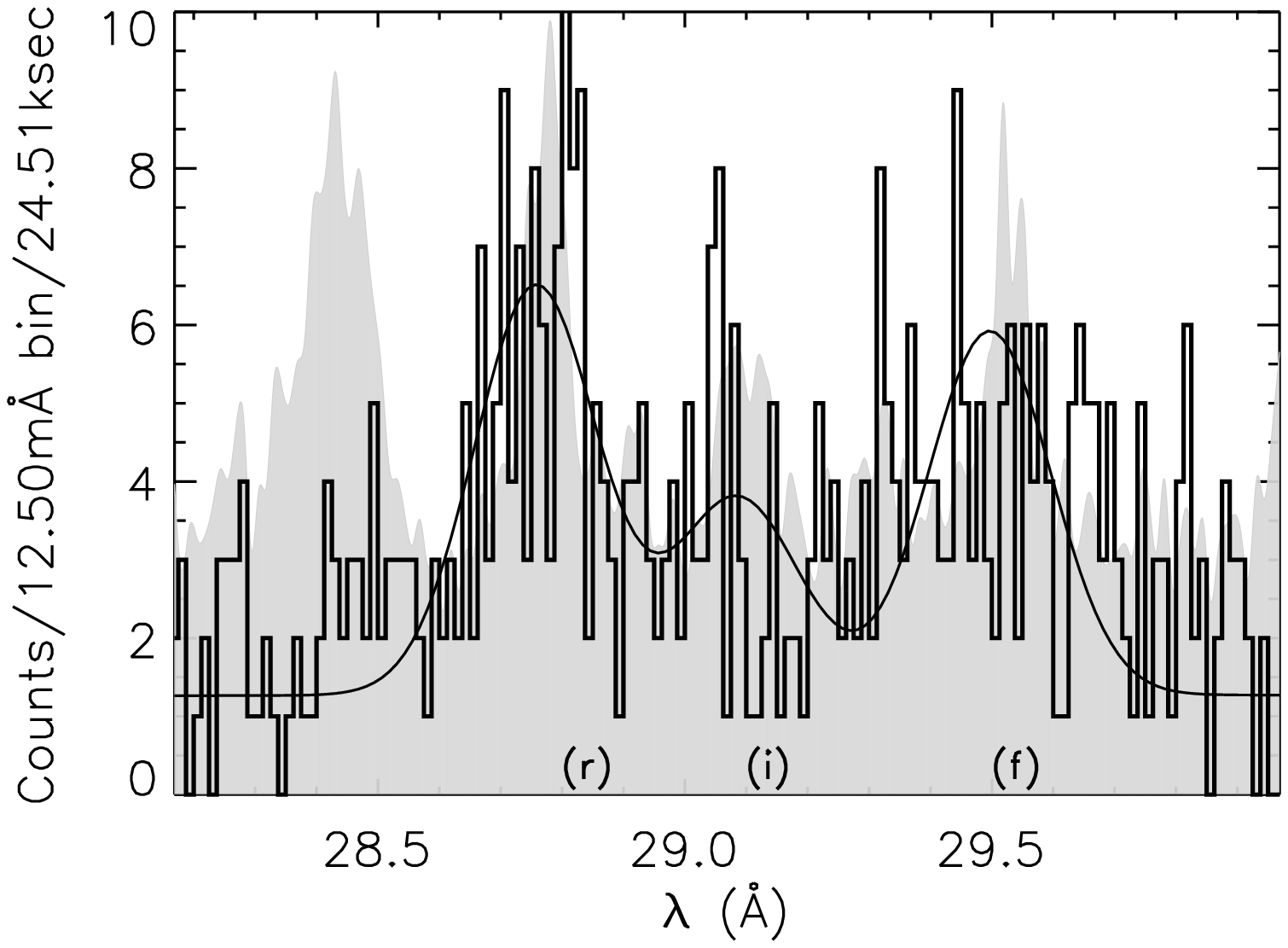}}
\caption{\label{he}O\,{\sc vii} (left) and N\,{\sc vi} (right) He-like triplets with
best-fitting one-component line templates. For O\,{\sc vii} no intercombination line
(expected at 21.8\,\AA) can be seen, while for the N\,{\sc vi} triplet this line
(expected at 29.1\,\AA) might be present. The intercombination lines are expected to
show up with increasing electron densities. The smooth filled curve is that of Capella.}
\end{figure*}

Apart from the lines of the He and H-like ions, it is difficult to make
definite identifications of the other lines. Most lines that occur in the region
from 30-50\,\AA\ are from transitions between excited states in Si, S, and Fe.
It is difficult to check the plausibility of possible identifications since
the $\Delta$n=0 resonance lines of the ions concerned lie at longer wavelengths
than can be observed with {\it Chandra}. We have used the mean apparent wavelength
shift and the mean line width
of the identified lines to predict the rest-frame wavelengths and values of line
widths of possible lines. Although the spectrum is noisy and some features are
narrower than expected, it is possible that some lines of Si\,{\sc xii} and
S\,{\sc xiv} are present. These are the S\,{\sc xiv} 3p-2s transitions at 30.43
and 30.47\,\AA\ and the Si\,{\sc xii} 3s-2p transitions at 45.52 and 45.69\,\AA.
The Si\,{\sc xii} 3p-2s transitions lie at $\sim 41$\,\AA\ and would not be
observable as they occur within a region of instrumental insensitivity.\\
For the lines listed in Table~\ref{ltab} we calculate the sum of the line
fluxes (corrected for N$_{\rm H}$) and find a total line luminosity of
$\sim 4\times10^{27}$\,erg\,s$^{-1}$.

\subsection{Densities}

From analyses of UV spectra \cite{shore03} reported values of hydrogen densities,
n$_{\rm H}$, between $1.25\times 10^7$ and $1.26\times 10^8$\,cm$^{-3}$. In the
X-ray regime no lines exist that can be used to measure n$_{\rm e}$ ($\approx$
n$_{\rm H}$) below $10^9$\,cm$^{-3}$. Since the X-ray lines are unlikely to be
found in the same region as the UV lines we checked the density-sensitive line
ratios in He\,{\sc i}-like ions \citep[e.g.,][]{gj69,denspaper}. Unfortunately
the signal to noise in the intersystem lines in N\,{\sc vi}, O\,{\sc vii} and
Ne\,{\sc ix} is too low to make quantitative analyses (see Figs.~\ref{lines} and
\ref{he}). Qualitatively, the upper limits to the intersystem lines suggest that
n$_e<2\times 10^9$\,cm$^{-3}$.

\subsection{Search for iron lines}
\label{fedef}

\begin{figure}
  \resizebox{\hsize}{!}{\includegraphics{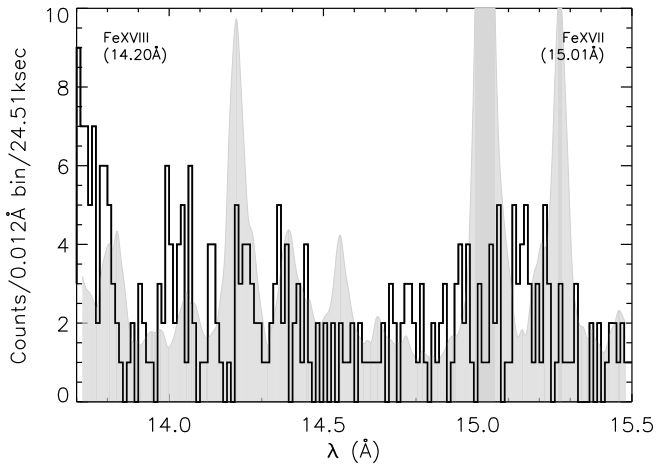}}
\caption{\label{felines}Spectral regions containing Fe\,{\sc xvii} expected at
15.01\,\AA, and Fe\,{\sc xviii} at 14.2\,\AA. No evidence for the presence of
these lines can be seen. For comparison the Capella LETG spectrum is
shown as smoothed filled curve.}
\end{figure}

From the bottom two panels of Fig.~\ref{lines} it can be seen that in addition
to the Ne lines identified in the spectrum of V382\,Vel, some lines appear in
the Capella spectrum that have no obvious counterparts in V382\,Vel. These lines
originate from Fe\,{\sc xvii} to Fe\,{\sc xxi} \citep[see][]{nebr}, and the absence
of these lines in V382\,Vel indicates an enhanced Ne/Fe abundance ratio. Since
V382\,Vel is an ONeMg nova, a significantly enhanced
Ne/Fe abundance (as compared to Capella) is expected. We have systematically
searched for line emission from the strongest Fe lines in different ionization stages.
We studied the emissivities of all iron lines below 55\,\AA\ predicted by the APEC
database for different stages of ionization as a function of temperature.
The strongest measureable line is expected to be that of Fe\,{\sc xvii} at 15.01\,\AA.
The spectral region around this line is shown in the upper left panel of
Fig.~\ref{felines}, and there is no evidence for the presence of this line.
The shaded spectrum is the arbitrarily scaled (as in Fig.~\ref{lines}) LETG spectrum
of Capella, showing where emission lines from Fe\,{\sc xvii} are expected.
There is some indication of line emission near 15.15\,\AA, which could be
O\,{\sc viii} Ly$_\gamma$ at 15.176\,\AA\ (Table~\ref{ltab}),
since the O\,{\sc viii} Ly$_\beta$ line at 16\,\AA\ is quite strong.
The spectral region shown in Fig.~\ref{felines} also contains the strongest
Fe\,{\sc xviii} line at 14.20\,\AA.
Again, in the Capella spectrum there is clear Fe\,{\sc xviii} emission, while
in V382\,Vel there is no indication of this line (a possible feature at
14.0\,\AA\ has no counterpart in the APEC database). Neither are the hotter
Fe\,{\sc xxiii} and Fe\,{\sc xxiv} lines detected in this spectrum. We conclude
that emission from Fe lines is not present in the spectrum.

\subsection{Temperature structure and abundances}
\label{temp}

\begin{figure*} 
  \resizebox{\hsize}{!}{\includegraphics{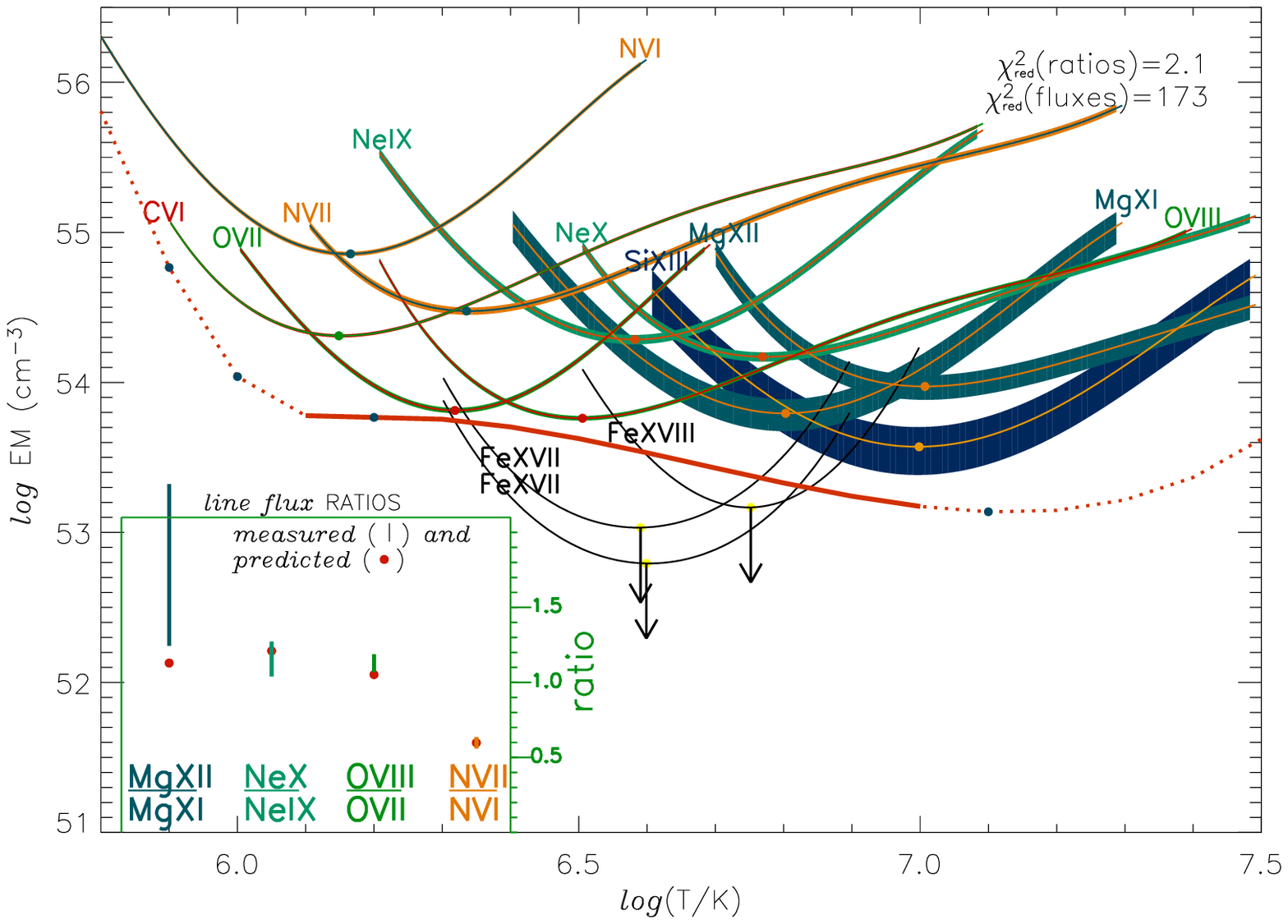}\includegraphics{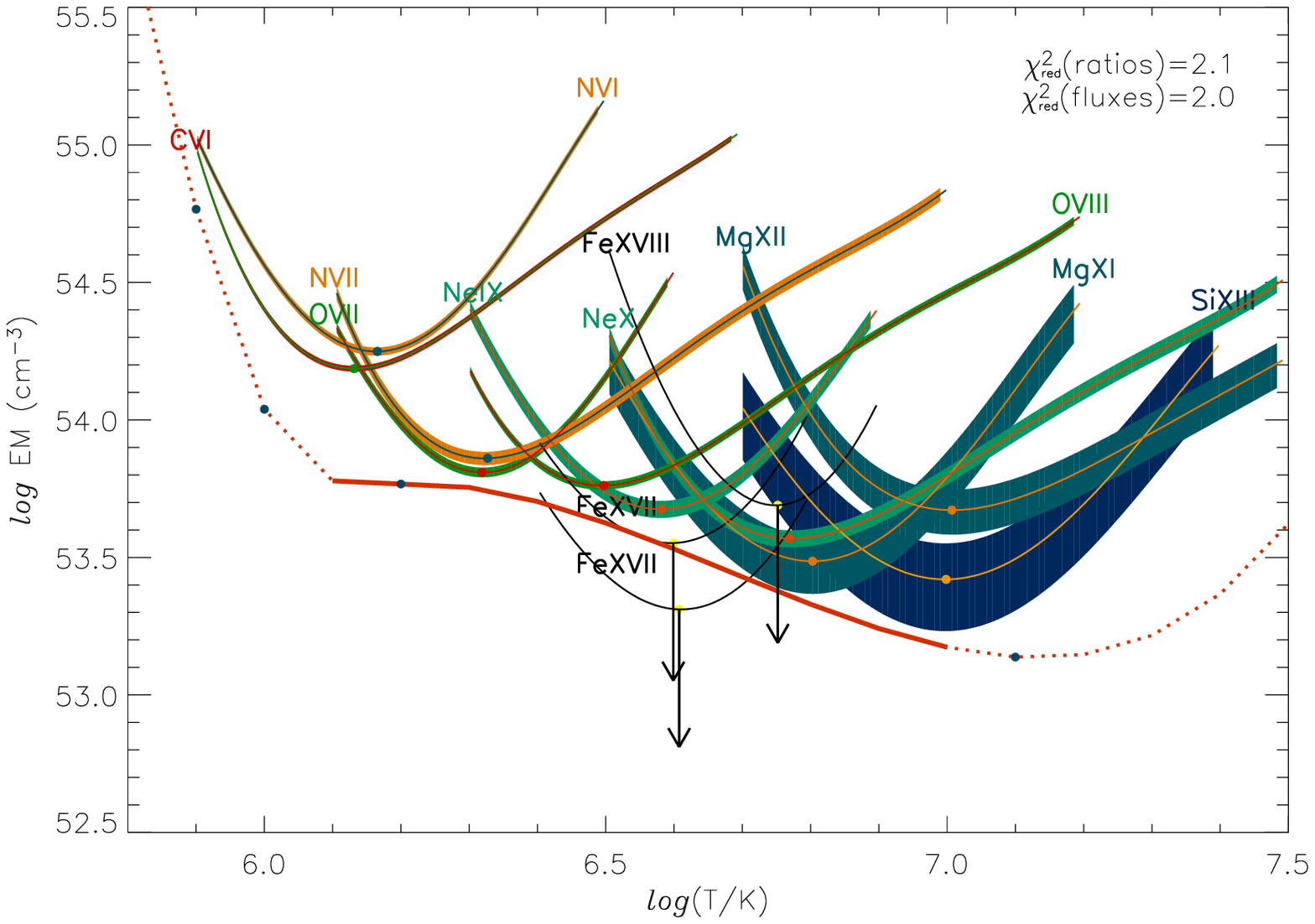}}
\caption{\label{loci}{\bf Left:} Emission measure loci of the strongest lines using
solar photospheric abundances.
The solid red curve marks the best fit reproducing the line ratios of Ly$_\alpha$
vs. He-like r line of Mg, Ne, O and N as well as the measured line fluxes of
O\,{\sc viii} and O\,{\sc vii}. The discrepancies in the loci of the other
lines are attributed to abundance effects. The loci for iron are upper limits.
The inset illustrates how well the indicated line ratios are reproduced.
{\bf Right}: The emission measure loci corrected by factors (Table~\ref{abund})
to give the corrections to the abundances of the respective elements
relative to O. The solid curve is the best fit reproducing the line fluxes.
Note that the y-axis has a different scale than in the left panel. The legends
give the $\chi^2$ values for the reproduction of the line ratios and of the line
fluxes.}
\end{figure*}

 The lines observed are formed under a range of temperature conditions. We assume
that collisional ionization dominates over photoionization and that the lines
are formed by collisional excitations and hence find the emission measure distribution
that can reproduce all the line fluxes. In Fig.~\ref{loci} we show emission
measure loci for 13 lines found from the emissivity curves $G_i(T)$
(for each line $i$ at temperature $T$) extracted from the Chianti database,
using the ionization balance by \cite{ar}. The volume emission measure loci are
obtained using the measured line fluxes $f_i$ (Table~\ref{ltab}) and
$EM_i(T)=4\pi\,{\rm d}^2(hc/\lambda_i)(f_i/G_i(T))$ having corrected the fluxes for N$_{\rm H}$.
The emissivities $G_i(T)$ are initially calculated assuming solar photospheric abundances \citep{asp}.
The solid smooth curve is the best-fitting emission measure distribution $EM(T)$, which can
be used to predict all line fluxes $F_i=hc/\lambda_i\int G_i(T)\,EM(T)dT(4\pi{\rm d}^2)^{-1}$.
In each iteration step the line fluxes for the
measured lines can be predicted ($F_i$) and, depending on the degree of agreement
with the measured fluxes, the curve parameters can be modified \citep[we used Powells
minimization -- see][]{numrec}. In order to exclude abundance effects at this stage we
optimized the reproduction of the temperature-sensitive line ratios of the Ly$_\alpha$
and He-like r lines of the same elements and calculated
the ratios $R_j=f_{{\rm Ly},j}/f_{{\rm r},j}$, for Mg, Ne, O, and N and then
compared these with the respective measured line ratios $r_j$
\citep[see][]{abun}.
This approach constrains the shape of the curve, but not the normalization, which is
obtained by additionally optimizing the O\,{\sc viii} and O\,{\sc vii} absolute
line fluxes; the emission measure distribution is thus normalized to the solar O
abundance. The goodness parameter is thus defined as
\begin{eqnarray}
\chi^2&=&\sum_j\frac{ (R_j-r_j)^2}{\Delta r_j^2}\\ \nonumber
&&+\,(f_{\mbox{O\,{\sc vii}}}-F_{\mbox{O\,{\sc vii}}})^2/\Delta f_{\mbox{O\,{\sc
vii}}}^2\\
&&+\,(f_{\mbox{O\,{\sc viii}}}-F_{\mbox{O\,{\sc viii}}})^2/\Delta f_{\mbox{O\,{\sc
viii}}}^2\nonumber
\end{eqnarray}
with $\Delta r$ and $\Delta f$ being the measurement errors of the indexed line ratios
and fluxes.
The emission measure distribution (solid red curve)
is represented by a spline interpolation between the interpolation points marked
by blue bullets. In the lower left we show the measured ratios (with error bars),
and the best-fitting ratios
(red bullets) showing that the fit has a high quality. We tested
different starting conditions (chosen by using different initial interpolation
points) and the fit found is stable.

 We used the best-fitting emission measure distribution to predict the fluxes
of the strongest lines in the spectrum and list the ratios of the measured and
predicted fluxes in Table~\ref{abund}. It can be seen that for a given element
the same trend is found for different ionization stages. If the mean emission
measure distribution predicts less flux than observed, then the element abundance
must be increased, to lower the loci to fit those of O.

Increased abundances (relative to O) are required for N and Ne (by a factor of
4), for Mg (by a factor of 2) and C and Si (by a factor of 1.4). The increase for
C, based only on the 33.8-\AA\ line is an upper limit, because the flux in this noisy
line may be overestimated. The Mg and Si line fluxes have quite large uncertainties
and the fitting procedure above $6\times 10^6$\,K is less secure. The resulting values
of N(N)/N(O) and N(Ne)/N(O) are 0.53 and 0.59, respectively, which are slightly lower
than those of \cite{shore03}, who found 0.63 and 0.79, respectively. The values of
N(Mg)/N(O) and N(Si)/N(O) are 0.19 and 0.1, respectively, which are substantially
larger than the values of 0.04 and 0.01 found by \cite{shore03}. To avoid too much
flux in the iron lines the value of N(Fe)/N(O) used must be reduced by $\sim 0.2-0.6$.
This gives N(Fe)/N(O)$<0.04$.

The corrected emission measure loci are shown in the right-hand panel of
Fig.~\ref{loci}, where it can be seen that the loci now form a smooth envelope
reflecting a meaningful physical distribution. The $\chi^2$ value in the upper right
corners of Fig.~\ref{loci} represent the goodness with which the ratios and the
O line fluxes are represented and how well all line fluxes (except iron) are
represented. We have tested the effects of changing the values of N$_{\rm H}$
by a factor of two. This primarily affects the longer wavelength lines
of C and N, leading to abundances (relative to O) that increase with
increasing values of the assumed N$_{\rm H}$. In Table~\ref{abund} we give examples
of the results for C and N using values of N$_{\rm H}$ that are twice the original
value of $1.2\times 10^{21}$\,cm$^{-2}$ and half of this value. The absolute emission
measure values increase by about 30 per cent when N$_{\rm H}$ is a factor of two higher
and decrease by about 20 per cent when N$_{\rm H}$ is lowered by a of factor two.
Without a measurement of the density we cannot derive a model from the absolute emission
measures. The values of the emission measures are consistent with those
given by \cite{orio04}.\\
We stress that only the average properties of the expanding
shell can be derived. Since we have evidence that the lines are produced
non-uniformly, there may be different abundances in different regions of the shell.

\begin{table}
\renewcommand{\arraystretch}{1.1}
\caption{\label{abund}Ratios of measured and predicted line fluxes from the best-fit
emission measure curve in the left panel of Fig.~\ref{loci}. These give the corrections
required to the adopted solar element abundances$^{[a]}$ \citep[][]{asp}, relative
to O.}
\begin{flushleft}
\begin{tabular}{lrlrlr}
\hline
\ \ ion & R$^{[b]}$\ \ & \ \ ion & R$^{[b]}$\ \ & \ \ ion & R$^{[b]}$\ \ \\
\hline
C\,{\sc vi}&    $<$1.3 & Si\,{\sc xiii}&  1.37  & Mg\,{\sc xii}  &  2.97 \\
N\,{\sc vii}&   3.92   & N\,{\sc vi}   &  3.92  & Mg\,{\sc xi}   &  1.88\\
O\,{\sc viii}&  0.98   & O\,{\sc vii}  &  0.91  & Fe\,{\sc xvii} &  $<$0.15\\
Ne\,{\sc x}&    3.79   & Ne\,{\sc ix}  &  3.97  & Fe\,{\sc xviii}&  $<$0.60\\
\hline
\multicolumn{6}{l}{with half the value of N$_{\rm H}$}\\
C\,{\sc vi}&    0.82   & N\,{\sc vi}   & 3.65   & N\,{\sc vii}&   3.66\\
\multicolumn{6}{l}{with double the value of N$_{\rm H}$}\\
C\,{\sc vi}&    3.38   & N\,{\sc vi}   & 4.48   & N\,{\sc vii}&   4.49\\
\hline
\end{tabular}
\\
$^{[a]}$ Adopted solar abundances (relative to O) C/O: 0.537, N/O: 0.132,
Ne/O: 0.151, Mg/O: 0.074, Si/O: 0.071, Fe/O: 0.062.\\
$^{[b]}$ Ratio of measured to predicted line fluxes.

\renewcommand{\arraystretch}{1}
\end{flushleft}
\end{table}

\section{Discussion}
\label{disc}

The previous X-ray observations of Nova V382\,Vel were carried out
at a lower spectral resolution making detection of line features
difficult. Both BeppoSAX and Chandra (ACIS-I) found that the nova
was extremely bright in the Super Soft Phase \citep{orio02,burw02}.
Orio et al. tried to fit their observations
with Non-LTE atmospheres characteristic of a hot WD with a `forest' of
unresolved absorption lines, but no reasonable fit was obtained.
These authors determined that even one or two unresolved emission lines
superimposed on the WD atmosphere could explain the spectrum,
and suggested that the observed `supersoft X-ray source' was
characterized by unresolved narrow emission lines
superimposed on the atmospheric continuum' \citep{orio02}. The
LETG spectrum obtained on 14 February 2000
shows emission lines with only a weak continuum. We conclude that nuclear burning
switched off before 2000 February and the emission peak at $\sim 0.5$\,keV reflects
continuum emission from nuclear burning.\\
The `afterglow' shows broad emission lines in the LETG spectrum reflecting the
velocity, temperature and abundance structure of the still expanding shell. The
X-ray data allow an independent determination of the absorbing column-density
from the ratio of the observed H-like Ly$_{\alpha}$ and Ly$_{\beta}$ line fluxes,
leading to a value consistent with
determinations in the UV. The value measured from the LETG spectrum appears to
represent the constant interstellar absorption value. Some of the lines
consist of several (at least two) components moving with
different velocities. These structured profiles
are different for different elements. While the O lines show quite compact
profiles, the Ne lines show a double feature indicative of two components.\\
No Fe lines could be detected although lines of Fe\,{\sc xvii} and Fe\,{\sc xviii}
are formed over the temperature range in which the other detected lines are formed.
We attribute this absence of iron lines to an under-abundance of Fe with respect to
that of O. Since we are observing nuclearly processed material from the white dwarf,
it is more likely that elements such as N, O, Ne, Mg and possibly C and Si are
over-abundant, rather than Fe being under-abundant. Unfortunately, no definitive
statements about the density of the X-ray emitting plasma can be made. Thus neither
the emitting volumes nor the radiative cooling time of the plasma can be found.

\section{Conclusions}
\label{conc}

We have analyzed the first high-dispersion spectrum of a classical
nova in outburst, but it was obtained after nuclear burning had ceased in
the surface layers of the WD. Our spectrum showed strong emission
lines from the ejected gas allowing us to determine velocities,
temperatures, and densities in this material. We also detect weak
continuum emission which we interpret as emission from the surface
of the WD, consistent with a black-body temperature
not exceeding $\sim 3 \times 10^5$\,K.
Our spectrum was taken only 6 weeks after an ACIS-S
spectrum that showed the nova was still in the Super Soft X-ray
phase of its outburst. These two observations show that the nova
not only turned off in 6 weeks but declined in radiated flux by a
factor of 200 over this time interval. This, therefore, is the
third nova for which a turn-off time has been determined and it
is, by far, the shortest such time. For example, ROSAT
observations of GQ\,Mus showed that it was declining for at least
two years if not longer \citep{shanley95} and ROSAT
observations of V1974\,Cyg showed that the decline took about 6
months \citep{krautt96}. Since the WD mass is not the only parameter,
but a fundamental one determining the turn-off time of the supersoft
X-rays, the mass of V382\,Vel may be larger than the two other novae 
\citep[see also][]{krautt96}.

The emission lines in our spectrum have broad profiles with FWHM
indicating ejection velocities exceeding 2900\,km\,s$^{-1}$.
However, lines from different ions exhibit different profiles. For
example, O\,{\sc viii}, 18.9\,\AA, and N\,{\sc vii}, 24.8\,\AA, can be fit by a
single Gaussian profile but Ne\,{\sc ix}, 13.45\,\AA, and Ne\,{\sc x}, 12.1\,\AA,
can only be fit with two Gaussians. We are then able to use the
emission measure distribution to derive relative element
abundances and find that Ne and N are significantly enriched with
respect to O. This result confirms that the X-ray regime is also
able to detect an ONeMg nova and strengthens the necessity of
further X-ray observations of novae at high dispersion.

\section*{Acknowledgments}

We thank the referee, Dr. M. Orio for useful comments that helped to improve the
paper. J.-U.N. acknowledges support from DLR under 50OR0105 and from PPARC under
grant number PPA/G/S/2003/00091. SS acknowledges partial support from NASA, NSF, and
CHANDRA grants to ASU.

\bibliographystyle{mn2e}
\bibliography{astron,jn,vel,cn}

\label{lastpage}

\end{document}